\documentclass[particles,review,accept,moreauthors,pdftex]{Definitions/mdpi} 
\usepackage{amssymb,color,graphicx,bm,mathrsfs,color,amsmath,slashed,comment}
\newcommand{\bea}{\begin{eqnarray}}
\newcommand{\eea}{\end{eqnarray}}
\newcommand{\be}{\begin{equation}}
\newcommand{\ee}{\end{equation}}

\input{acronym.input}
\definecolor{red}{rgb}{0.8,0,0}
\definecolor{violet}{rgb}{0.4,0,0.4}
\definecolor{green}{rgb}{0,0.5,0.0}
\definecolor{navy}{rgb}{0.0,0.0,0.6}
\definecolor{orange}{rgb}{0.8,0.2,0.0}

\usepackage[normalem]{ulem}  

\firstpage{1} 
\doinum{https://doi.org/10.3390/particles5040037}
\pubvolume{5}
\issuenum{4}
\articlenumber{493-513}
\pubyear{2022}
\copyrightyear{2022}
\externaleditor{Academic Editor: Armen Sedrakian}
\history{Received: 9 August 2022; Accepted: 30 August 2022; Published: 17 November 2022}
\updates{yes}

\Title{Formation, Possible Detection and Consequences of Highly Magnetized Compact Stars} 

\Author{
	Banibrata Mukhopadhyay $^{1,*,\dagger}$\orcidB{} 
	and Mukul Bhattacharya $^{2,*,\dagger}$\orcidA{} 
            }

\address{
\quad \textls[-25]{${}^1$ Department of Physics, Indian Institute of Science, Bangalore 560012, India} \\  
\quad \textls[-25]{${}^2$ Department of Physics, The Pennsylvania State University, University Park, PA 16802, USA
} 
} 
\corres{Correspondence: bm@iisc.ac.in (B.M.);  mmb5946@psu.edu (M.B.); Tel.: +91-80-2360-7704 (B.M.); Tel.: +1-512-293-7935 (M.B.); }
 
\firstnote{These authors contributed equally to this work.} 
 
\AuthorNames{Mukul Bhattacharya and Banibrata Mukhopadhyay}

\abstract{Over the past several years, there has been enormous interest in massive neutron stars and white dwarfs due to either their direct or indirect evidence. The recent detection of gravitational wave event GW190814 has confirmed the existence of compact stars with masses as high as $\sim$2.5--2.67$M_{\odot}$ within the so-called mass gap, indicating the existence of highly massive neutron stars. One of the primary goals to invoke massive compact objects was to explain the recent detections of over a dozen Type Ia supernovae, whose peculiarity lies with their unusual light curve, in particular the high luminosity and low ejecta velocity. In a series of recent papers, our group has proposed that highly magnetised white dwarfs with super-Chandrasekhar masses can be promising candidates for the progenitors of these peculiar supernovae. The mass-radius relations of these magnetised stars are significantly different from those of their non-magnetised counterparts, which leads to a revised super-Chandrasekhar mass-limit. These compact stars have wider ranging implications, including those for soft gamma-ray repeaters, anomalous X-ray pulsars, white dwarf pulsars and gravitational radiation. Here we 
review the development of the subject over the last decade or so, 
describing the overall state of the art of the subject as it stands now.
We mainly touch upon the possible formation channels of these intriguing stars as well as the effectiveness of direct detection methods. These magnetised stars can have many interesting consequences, including reconsideration of them as possible standard candles.}


\begin{document}

\section{Introduction}
\label{Sec1}

In the last several years, there has been much interest in massive compact objects.
Sometimes the evidence is direct; however, it is indirect on some other occasions.
For instance, the~detection of gravitational wave (GW) event GW190814 \cite{Abbott2020} directly confirms the existence of a compact star with a mass of 2.5--2.67$M_\odot$. Thus far, the minimum mass of an astrophysical black hole is understood to be around 
 $3M_\odot$ \cite{mac,pooley} (also see~\cite{ozel}) while the maximum mass of a neutron star (NS) is argued to be about $2.5M_\odot$ \cite{alsing}, leading to a so-called mass gap.
Therefore, the~above observation fills in the gap, arguing the compact
object to be a massive NS. 
Note, however, that some researchers also claim against this mass gap.
There indeed can be a depression in mass distribution.
Nevertheless, there is other evidence for massive
NSs (although not as massive as that inferred from GW190814) based
on pulsar observations~\cite{Cromartie2020} with mass $>2M_\odot$. 
Indeed, the statistics based on observation showed that while the averaged mass of NSs is about $1.4M_\odot$, the~accreted millisecond pulsars have masses, on~average, of $1.6M_\odot$ \cite{pul,miller}. Hence, there is no surprise with the discovery of more than two solar mass pulsars.

On the other hand, for~more than
15 years, at~least a dozen objects have been detected as peculiar over-luminous
Type Ia Supernovae (SNeIa), whose unusually high luminosity, along with their violation of
Khokhlov's limit~\cite{khok}, argue for progenitor masses that are as high as $2.8M_\odot$
\cite{Scalzo2010,Howell2006}.
This is evidence for a significant violation of the Chandrasekhar mass-limit of
$\sim$1.4$M_\odot$ of white dwarfs (WDs) and a super-Chandrasekhar~mass-limit.

An important question to ask is: how do we argue for the existence of such unconventional and massive compact
objects? Numerous investigations have been carried out by our group aimed at uncovering this mystery over the last one decade. 
We have shown that strong magnetic field: its strength and anisotropic
effect~\cite{DM2012,DM2013,DM2013b,DM2014,DM2015,SM2015,MB2018,MB_euro2018,AG2020,MB2021,MB2022}, modified Einstein's gravity~\cite{Ban2017,EPL2019}, and matter encountering noncommutative physics at high density in the core of compact stars~\cite{IJMPD2020,IJMPD2021} can each independently lead to highly massive WDs and NSs.
After our initiation, several other groups independently started looking
at this issue based on other ideas, e.g., the~ungravity effect~\cite{BertM2016}, WDs
having net electric charge~\cite{Liu2014,Car2018}, lepton number violation~\cite{Belyaev2015} and anisotropic pressure~\cite{HB2013}.
Additionally, there appears to be a relation between equation of state (EoS) 
and compact mass~\cite{glen}.

Magnetic fields in WDs and NSs, in~general, degenerate stars, have been 
explored for a long time (see the review~\cite{cham} from three decades ago).
In general, it is not difficult to understand the surface field of WDs and
NSs to be $\sim$$10^9$ and $10^{15}$ G, respectively~\cite{cham,fw2005}, 
based on their different origins, e.g.,~fossil effect, dynamo, binary
evolution etc.~\cite{fw2000,fwk}. Spectropolarimetric observations suggest that the magnetic 
flux of majority magnetic WDs is similar to that of main-sequence Ap-Bp 
stars~\cite{fw2005}. This argues for an evolutionary link of
magnetism between the main sequence stars and WDs. A similar argument
can be made for isolated NSs if the underlying O-type progenitors have
effective dipolar fields $\sim$$10^3$ G~\cite{fw2005}. It has been statistically found
that magnetic WDs are, in general, massive, and their number with fields
at high as $\sim$$10^9$ G could even be $10\%$ of the total population~\cite{fwk}. However, magnetic field evolutions for isolated and accreting 
WDs are different. This is no surprise because~in the outer layers
of WDs, the field structure is expected to be modified due to accretion,
particularly above a critical accretion rate \mbox{rate $\approx \text{(1--5)}\times 
10^{-10} M_\odot {\rm yr}^{-1}$}. This results in a shorter Ohmic decay time,
leading to an apparent lack of field $\sim$$10^9$ G in observed accreting WDs~\cite{Cumming2002,zwf}.

Among other consequences of magnetized and modified gravity-induced WDs
are WD pulsars. These stars can also behave as soft gamma-ray repeaters (SGRs)
and anomalous X-ray pulsars (AXPs) \cite{MR2016,BM2016,BM2019}, generating significant gravitational
radiation that can be detected by space-based detectors~\cite{Kalita2019,Kalita2020,Kalita2021}. 
Many of these transients can eventually turn out to be WDs with super-Chandrasekhar masses. 
Observational data from the Sloan Digital Sky Survey (SDSS) suggest that 
magnetic WDs can have larger masses compared to their non-magnetic counterparts, although~they span the same temperature range~\cite{Van2005,Ferrario2015}. Regardless of the rotation rate, strong magnetic fields have been shown to modify the EoS of electron degenerate matter and yield super-Chandrasekhar WDs with masses up to $\sim$$2.6\, M_{\odot}$ \cite{DM2012,DM2013,DM2013b,SM2015}. 

The effect of strong magnetic fields on the mass-radius relation has been explored previously by our group for~both Newtonian~\cite{DM2012} and general relativistic~\cite{DM2014,SM2015,DM2015,MB2021} formalisms. These studies were performed for various different field configurations and were in good agreement with the results from independent studies~\cite{Bos2013,FS2015,Car2018}. All the above ventures, including the underlying observations, by~the aforementioned groups have brought super-Chandrasekhar WDs into the limelight in recent times. Magnetized WDs (B-WDs) have many important implications other than their apparent link to peculiar SNeIa, and, therefore, their other properties should also be explored~\cite{MR2016,BM2016}. Strong magnetic fields can influence the thermal properties of WDs, namely their luminosity, temperature gradient and cooling rate~\cite{MB2018,AG2020, MB2021,MB2022}. 
Moreover, the~magnetic field of WD has a crucial role in their binary 
evolution as progenitors of SNeIa or accretion-induced collapse~\cite{imin1,imin2,imin3}. Additionally, a possible merger through 
binary evolution leading to a peculiar SNeIa could be an 
alternate explanation for super-Chandrasekhar progenitors~\cite{imin4}.

In this review paper, we discuss the broad implications of these compact stars as well as the current status of observations. We aim at gathering all the 
underlying results obtained in last decade or so and try to assess the 
overall state of the art of the subject as it stands now, in~particular regarding the effects of magnetic
fields in NS and WD. This paper is organised as follows. In~Section~\ref{Sec2}, we discuss the origin and evolution of strong magnetic fields inside WD stars. In~Section~\ref{Sec3}, we simulate the rotating and magnetized B-WDs based 
on their structural stability for toroidal and poloidal magnetic field geometries. In~Section~\ref{Sec4}, we compute the mass-radius relation for non-rotating B-WDs, assuming a spherical geometry and radially varying field strength. In~Section~\ref{Sec5}, we explore the effect of cooling as well as magnetic field dissipation on the modified mass limit and the suppressed luminosities of B-WDs. In~Section~\ref{Sec6}, we briefly describe the quantisation of energy states of 
matter in B-WDs and derive their modified mass limit, assuming both constant and varying magnetic fields. In~Section~\ref{Sec7}, we investigate the GW emission originating from WDs with misaligned rotation and magnetic axes, along with their detection prospects.  Finally, we discuss the effect of anisotropy of dense matter in~the presence of strong magnetic fields on~the properties of B-WDs in Section~\ref{Sec8} and conclude in Section~\ref{Sec9}.

\section{Origin and Evolution of Magnetic Fields inside White~Dwarfs}
\label{Sec2}
Here, we discuss the origin of strong fields inside magnetised WDs and their evolution at the end of the main sequence during~the asymptotic giant phase. We present both numerical results from the Cambridge stellar evolution code STARS and analytical estimates from virial theorem that provide a relation between various energy components of the~WD.

\subsection{Origin of Magnetic~Field}
Magnetic fields in stars exhibit complex behaviour with structure on both small and large scales. The~origin of large-scale fields is debated. They can be fossil fields (e.g.,~\cite{DM2013b,fw2005}) that can be dated back to the formation of the star or~fields generated by a dynamic effect. Additionally, the field can be generated due to WD mergers (e.g.,
~\cite{fw2005}). As large-scale magnetic fields in stars can be unstable, dynamo effects are interesting because they can regenerate these fields. 
It is well known that purely poloidal or toroidal fields in stars are both structurally unstable~\cite{MT1973,Tayler1973}. However, magnetized stars and B-WDs with a toroidally dominated mixed field configuration, along with a small poloidal component, are among the most plausible cases~\cite{Wick2013} and also have an approximately spherical shape~\cite{SM2015}. Unlike the surface magnetic field, which can be observationally inferred, fields in the interior of the WD cannot be directly constrained. However, there is a sufficient evidence that stars exhibit dipolar fields in their outer regions and are expected to
have stronger toroidally dominated fields in their interior. Numerical simulations have already shown that the central fields of B-WDs can be several orders of magnitude higher than those at the surface~\cite{SM2015,BM2017,QT2018}.

\subsubsection{Dynamo-Model Based STARS~Simulations}
\begin{figure}[H]%
\begin{center}
	\parbox{2.75in}{\includegraphics[width=2.75in]{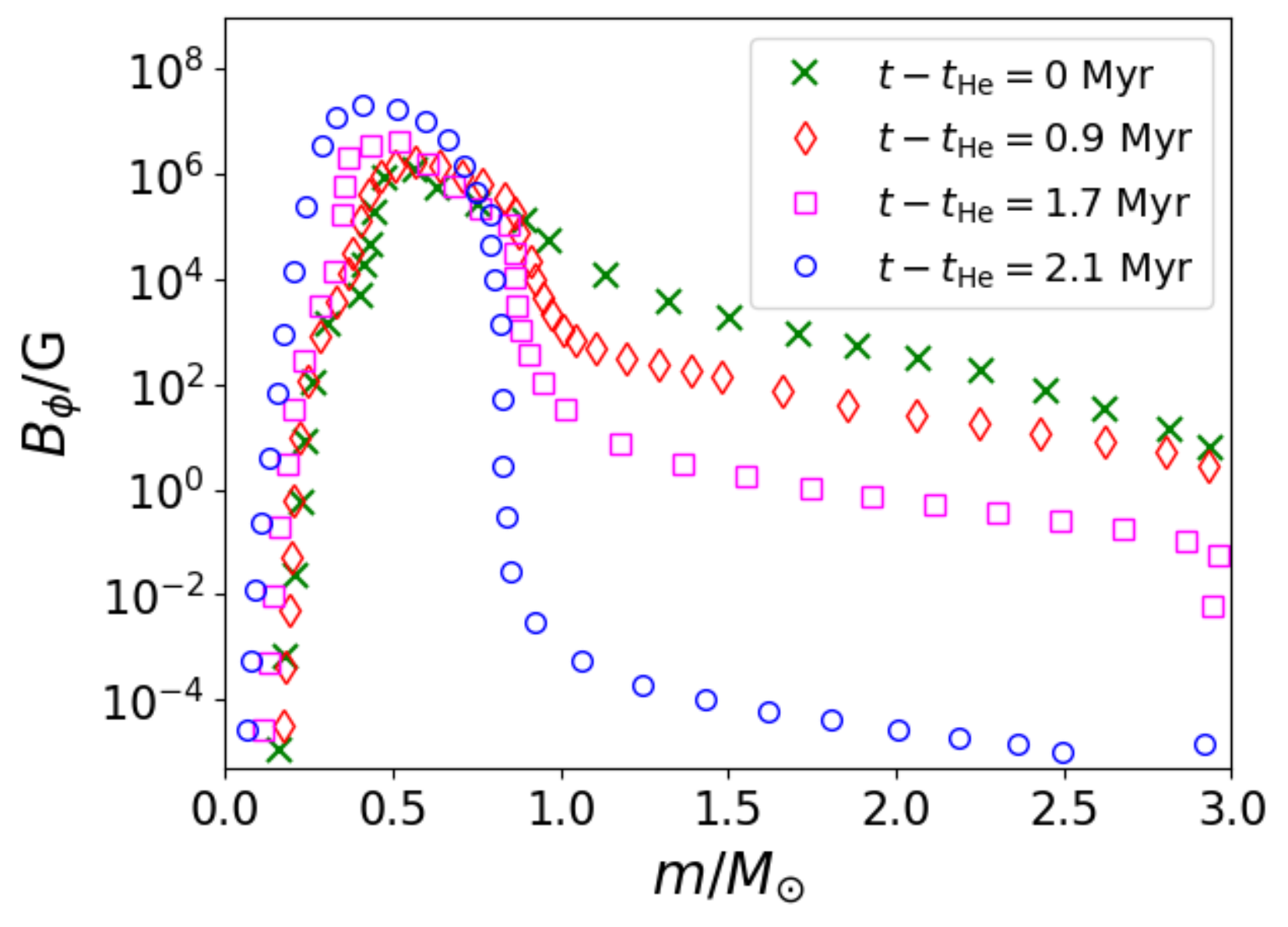}} 
	\hspace*{1pt}
  \caption{ 
	Toroidal field inside the star is shown as a function of the mass coordinate at various times after 
	the helium exhaustion in the core, during the asymptotic giant phase.
	See \cite{QT2018} for details.}%
 \label{fig1}
 \end{center}
\end{figure}

The evolution of magnetic field components along with the angular momentum 
during stellar evolution has been modelled recently with the Cambridge stellar evolution code STARS~\cite{QT2018} using advection-diffusion equations coupled to the structural and compositional equations. It has been shown that the magnetic field is likely to be dipolar in nature, decaying as an inverse square law for most of the star. The~simulation results also suggest that, at~the end of main sequence, stars have toroidally dominated magnetic fields. 
The authors reported \cite{QT2018} the evolution of toroidal field in the interior of the WD as a function of mass coordinate at various times in the end of main sequence after the exhaustion of helium nuclei in the core, i.e. during the asymptotic giant phase, as displayed in Figure \ref{fig1} for completeness.


Large-scale magnetic fields can be present in the degenerate core of magnetised WDs or B-WDs; even during the late stages of stellar evolution~\cite{QT2018} very high fields can develop based on the conservation of magnetic flux, as well as~from the dynamo mechanism. Therefore, strong fields inside magnetized WDs can also be of fossil origin. While the mass of the WD increases due to accretion, the magnetic field is advected into its interior. Consequently, the~gravitational power dominates over the degeneracy pressure, which leads to the contraction of the star. Hence, the~initial seed magnetic field is amplified as the total magnetic flux remains~conserved.

For magnetic field $B \sim 10^8\, {\rm G}$ within a star of size $R \sim 10^6\, {\rm km}$, the resultant flux will be $\sim$$10^{20}\, {\rm G\, km^2}$. From~flux freezing, for~a 1000 km size B-WD, the~magnetic field can then grow up to $\sim$$10^{14}\, {\rm G}$. Once the field increases, the~total outward force further builds up to balance the inward gravitational force, and the whole cycle is repeated multiple times. Therefore, the~magnetic fields of B-WDs are likely to be fossil remnants from their main-sequence progenitor~stars.

\subsubsection{Modified Virial~Theorem}
Virial theorem relates the integrated gravitational potential, thermal, kinetic and magnetic energies of a physical system to provide insight into its equilibrium configuration. By~invoking magnetic flux conservation and based on the
variation of the internal magnetic field with the matter density as a power law, the~modified virial theorem can be derived using the equation of magnetostatic equilibrium~\cite{AS2021}. The~well-known form of the virial theorem is
\begin{equation}
2T + W + 3\Pi + \mu = 0,
\end{equation}
where $T$, $W$, $\Pi$ and $\mu$ are the kinetic, gravitational, thermal and magnetic energies, respectively. For~the case of a static non-rotating WD, we have $T=0$. Assuming that a polytropic EoS is satisfied through the entire star such
that $P = K\rho^\Gamma$, where $K$ and $\Gamma$ are polytropic constants, and~$M=(4\pi/3)R^3 \rho$ with $\rho$ is the mean density, the~scalar virial theorem reduces to
\begin{equation}
-\alpha \frac{GM^2}{R} + \beta \frac{M^\Gamma}{R^{3(\Gamma-1)}} + \gamma \frac{\Phi_M^2}{R} = 0.
\end{equation}

After rearranging the terms, we obtain
\begin{equation}
M = \sqrt{\frac{\gamma \Phi_M^2}{\alpha G \left(1 - \frac{\beta M^{\Gamma-2}}{\alpha G R^{3\Gamma-4}}\right)}}
\end{equation}
for any $\Gamma$. For~$\Gamma = 4/3$, appropriate for extremely relativistic non-magnetic degenerate electrons, we obtain a mass which is independent of radius for a fixed magnetic flux, as~expected from Chandrasekhar's theory.  We evaluate the coefficients $\alpha$, $\beta$ and $\gamma$ to establish the modified virial theorem for a high magnetic~field. 

In the presence of a strong magnetic field, the~contribution of the magnetic pressure to the magnetostatic balance cannot be neglected. The~new momentum balance condition, neglecting the effect of magnetic tension, is then given by
\begin{equation}
\frac{1}{\rho}\left(\frac{dP}{dr} + \frac{dP_B}{dr}\right) = - \frac{Gm(r)}{r^2}
\end{equation}
at an arbitrary radius $r$ with mass enclosed at that radius $m(r)$, where $\rho$ includes the contribution from magnetic field and $P_B$ is the pressure due to the magnetic field of the star. We consider two different approaches to compute the modified virial theorem (see~\cite{AS2021}): (i) invoke flux conservation (freezing), which is quite common in stars when conductivity is high, and~(ii) assume 
the magnetic field varies as a power law with respect to density, just as the EoS of thermal pressure, throughout. 

\begin{figure}[H]
\begin{tabular}{cc}
\includegraphics[width=2.56in]{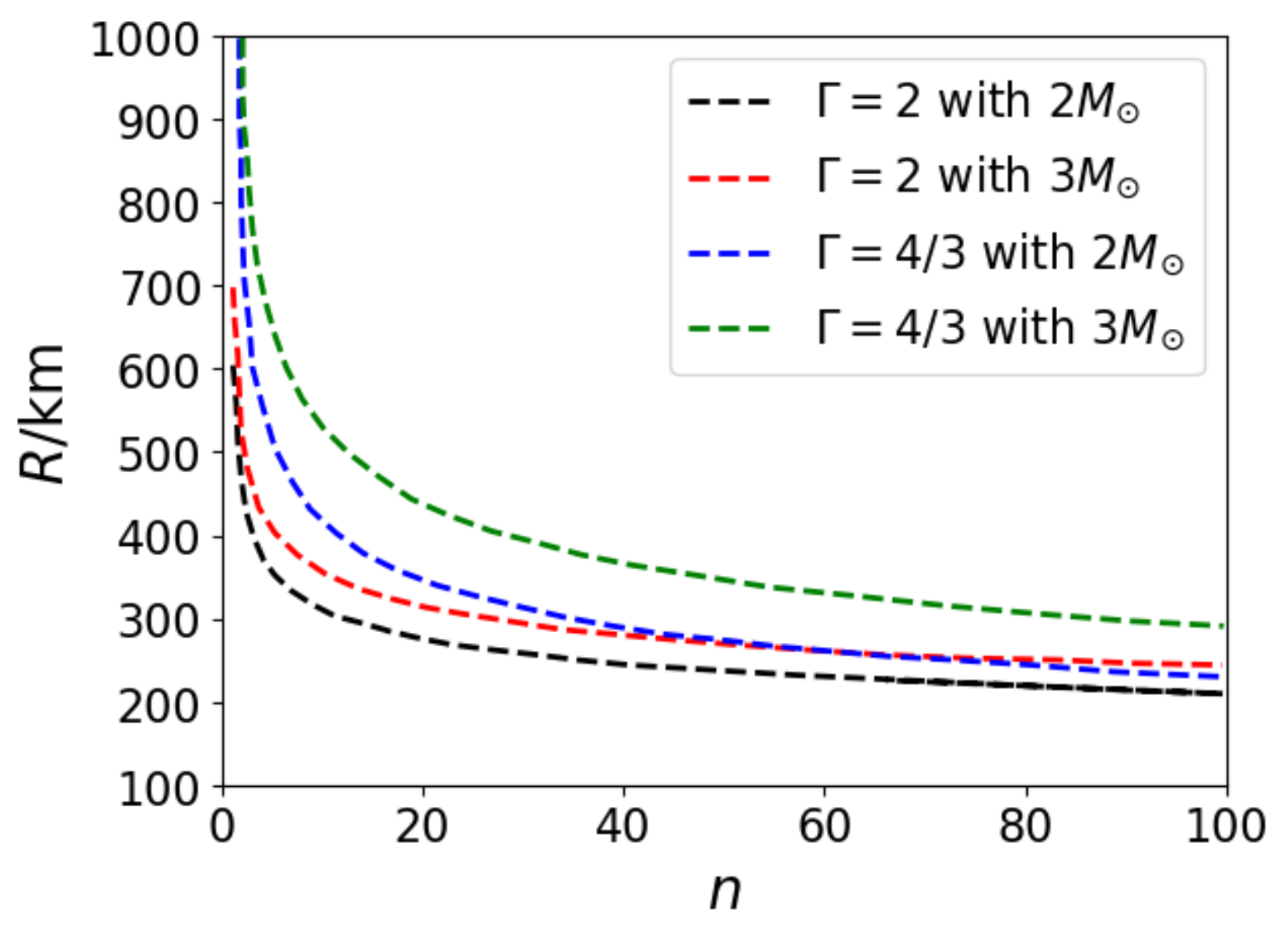} 
&\includegraphics[width=2.56in]{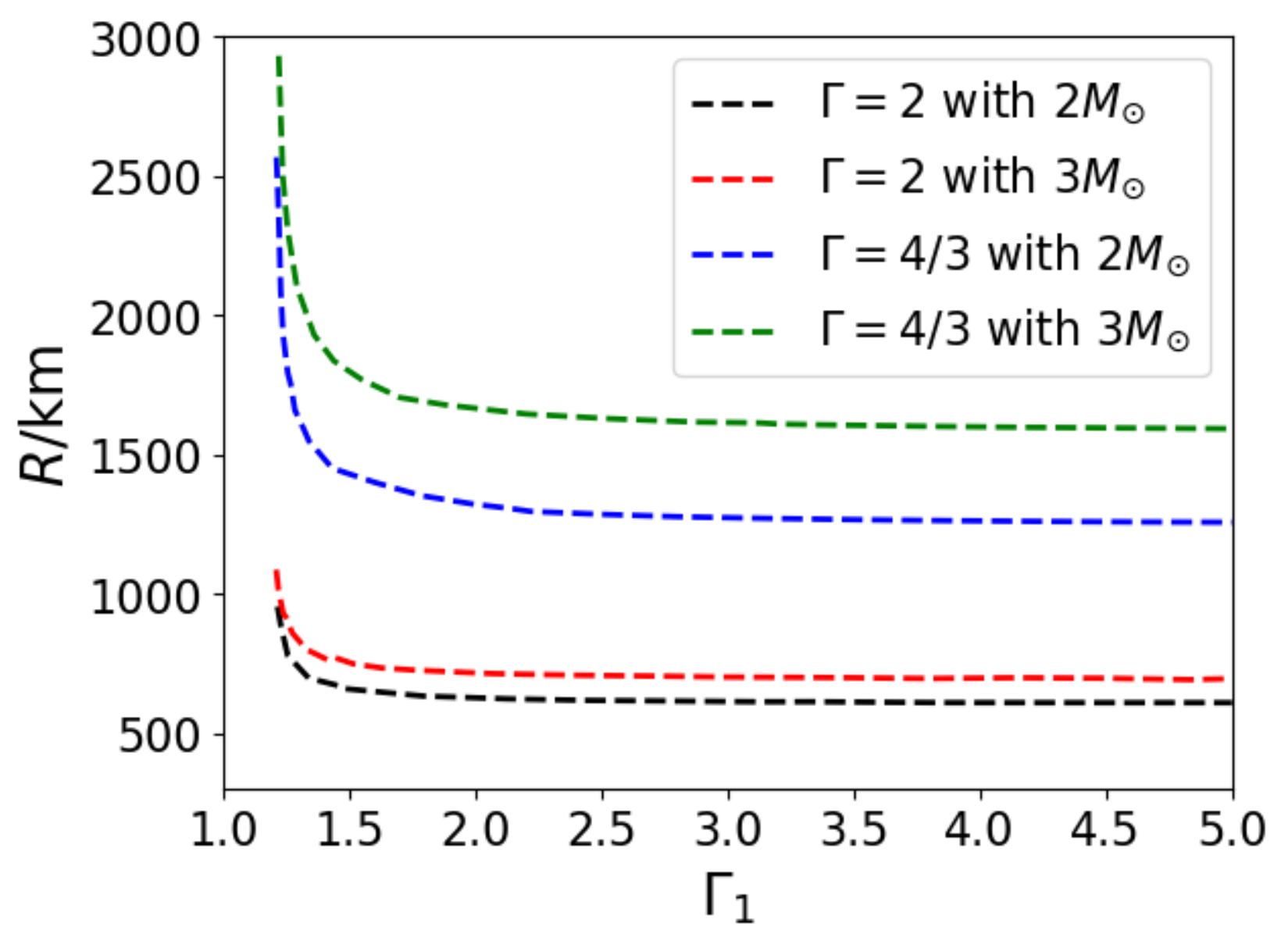}\\ 
\end{tabular}
  \caption{ 
	{\it Left panel:} Variation of radius with $n$ for the conserved flux model with various $\Gamma$ and total masses. 
	{\it Right panel:} Variation of radius with $\Gamma_1$ for the power-law field model with various $\Gamma$ and total masses.
	In each case, $\Gamma = 4/3$ corresponds to $\overline{B} = 10^{14}\ {\rm G}$, and $\Gamma= 2$ corresponds to $\overline{B} = 10^{16}\ {\rm G}$, with
	$\overline{B}$ being the average magnetic field. Parameter sets from 
	top to bottom mentioned in the figures correspond to the lines 
	sequentially from bottom to top in the low $n$ and $\Gamma_1$ regimes. See~\cite{AS2021}.}%
 \label{fig2}
\end{figure}

For the case (i), the~coefficients of the modified virial theorem are obtained as
\begin{equation}
\alpha = \frac{3(\Gamma-1)}{5\Gamma-6},\ \beta = \frac{3^\Gamma K}{(4\pi)^{\Gamma-1}},\ \gamma = \frac{2(n-1)}{5\Gamma-6} + \frac{4n-3}{6}.
\end{equation}
For $n=1$, the situation simplifies to that of a non-magnetized or weakly magnetized WD or a B-WD with constant magnetic field and hence constant $P_B$ throughout. For~power-law fields of case (ii), corresponding coefficients are
\begin{equation}
\alpha = \frac{3(\Gamma_1-1)}{5\Gamma_1-6},\ \beta = \left(1 + \frac{\Gamma - \Gamma_1}{(5\Gamma_1 - 6)(\Gamma-1)}\right)\frac{3^\Gamma K}{(4\pi)^{\Gamma-1}},\ \gamma = \frac{1}{6},
\end{equation}
where $\Gamma_1$ is the constant from the relation $P_B = K_1 \rho^{\Gamma_1}$. It is important to note that $\alpha$ is related to the scaling of $B$ with $\rho$. Therefore, the~presence of magnetic pressure allows either a more massive or smaller star. For~$\Gamma=\Gamma_1$, the~result reduces to that of the non-magnetic case with a redefined $K$. Figure~\ref{fig2} shows the variation of radius with indices $n$ and $\Gamma_1$ for the conserved flux and power-law field models, respectively, with~various $\Gamma$ and total~masses.

\subsection{Evolution of Magnetic~Field}
Highly magnetised WDs can possibly result from repeated episodes of accretion and spin-down~\cite{BM2017}. The~entire evolution of the B-WD can be classified in two phases: accretion-powered and rotation-powered. The~accretion-powered phase is governed by three conservation laws: linear and angular momenta conservation and conservation of magnetic flux around the stellar surface, given as
\begin{equation}
l\Omega(t)^2 R(t) = \frac{GM(t)}{R(t)^2},\ I(t)\Omega(t) = {\rm constant},\ B_s(t)R(t)^2 = {\rm constant},
\end{equation}
where $l>1$ accounts for the dominance of gravitational force over the centrifugal force, $I$ is the stellar moment of inertia and $\Omega$ is the angular velocity of the star that also includes the contribution acquired due to accretion. Here, we neglect any possible field decay and assume a specific field 
geometry. Solving the above equations simultaneously gives the time evolutions of the radius, magnetic field and angular velocity during the accretion phase. Accretion discontinues when
\begin{equation}
-\frac{GM}{R^2} = \frac{1}{\rho} \frac{d}{dr} \left(\frac{B^2}{8\pi}\right)_{r=R} \sim -\frac{B_s^2}{8\pi R\rho},
\end{equation}
where $\rho$ is the density of the inner disk~edge.

\vspace{-3pt}
\begin{figure}[H]
\begin{tabular}{cc}
\includegraphics[width=2.35in]{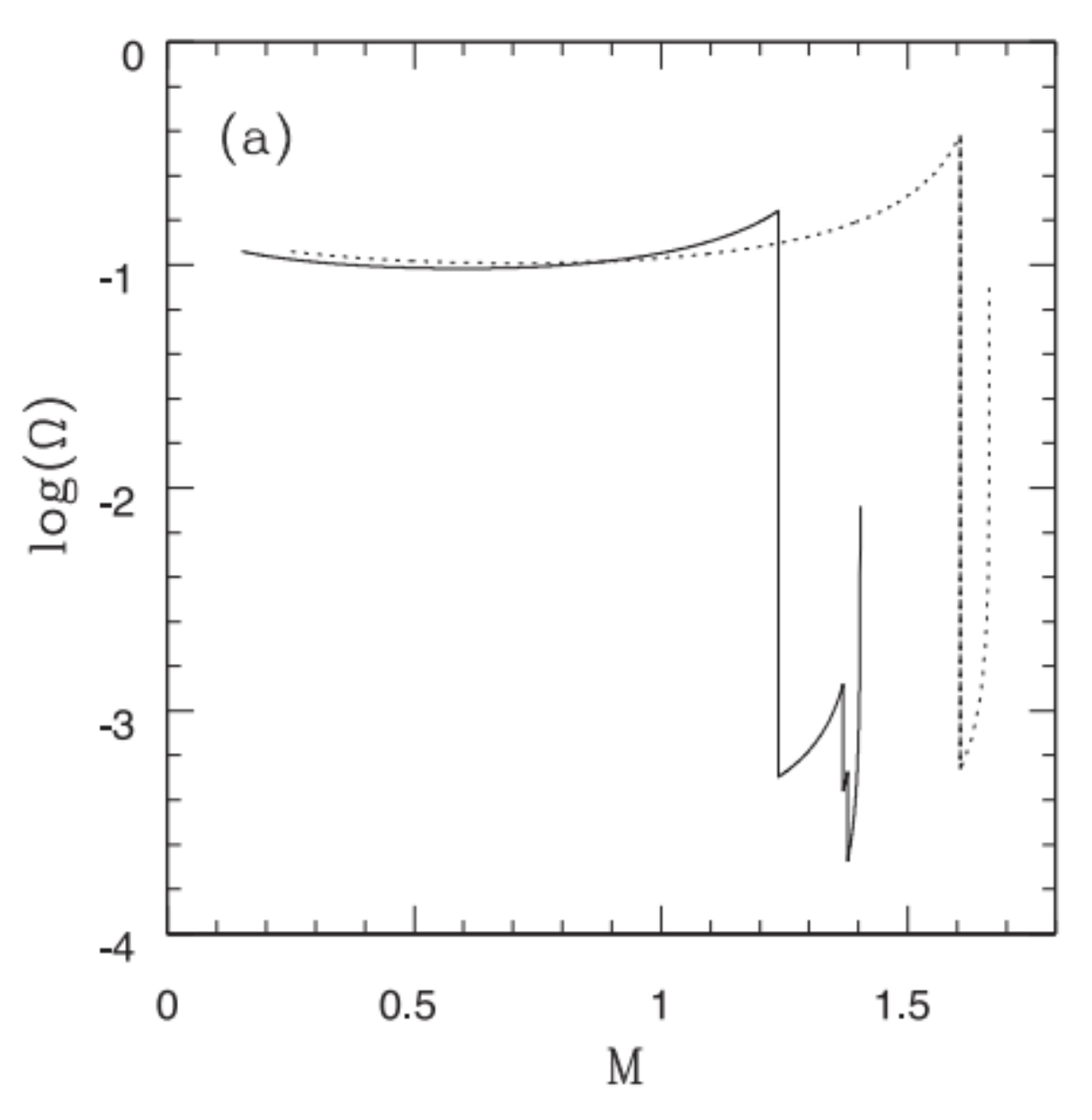}
&\includegraphics[width=2.35in]{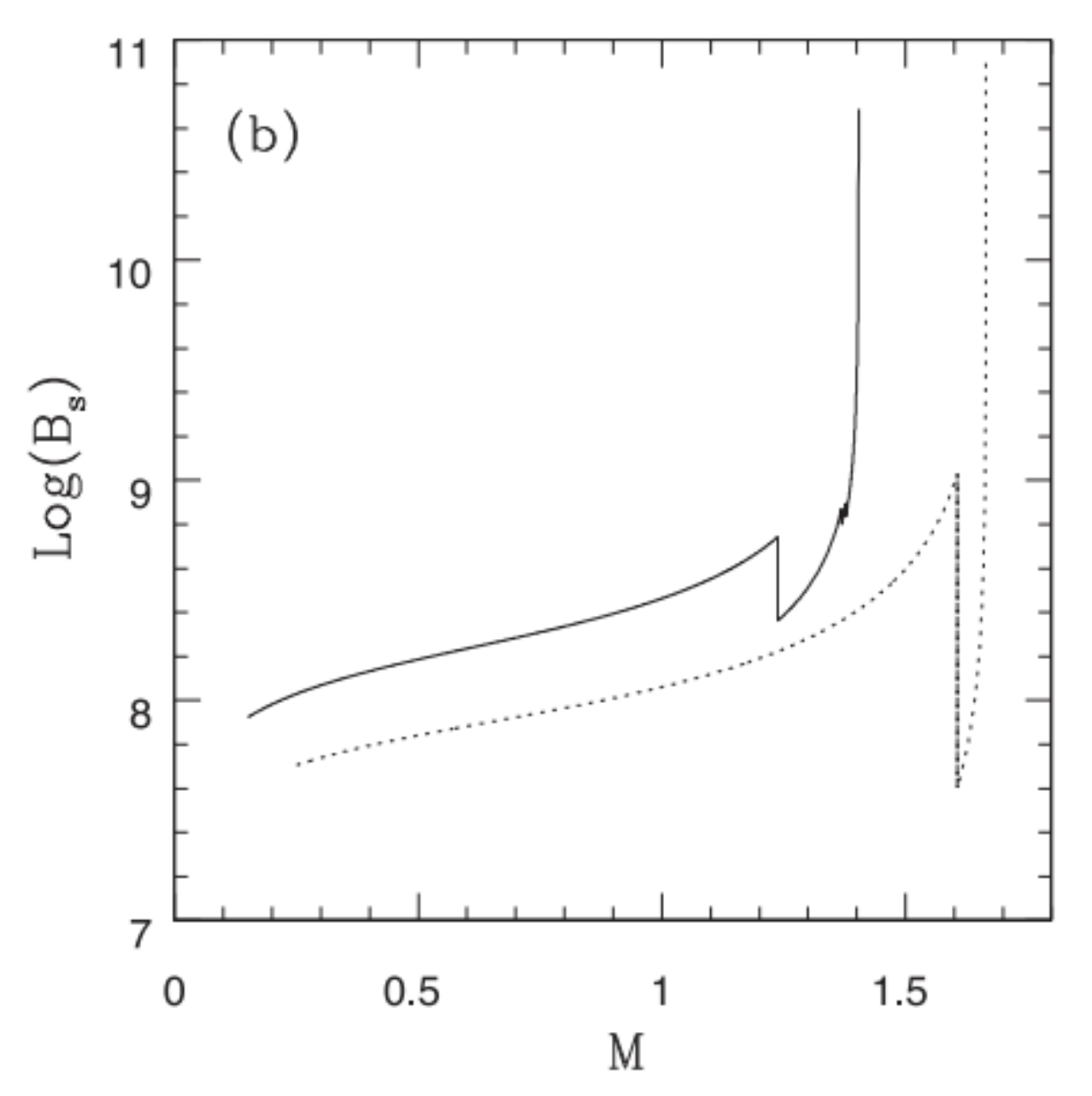}\\
\end{tabular}
	\caption{Time 
 evolution of angular velocity (left panel) and magnetic field (right panel) as functions of mass. The~solid curves correspond to the case with $n=3$, $m=2.7$, $\rho=0.05\, {\rm g\, cm^{-3}}$, $l=1.5$ and dotted curves correspond to the case with $n=3$, $m=2$, $\rho=0.1\, {\rm g\, cm^{-3}}$, and $l=2.5$. The~other parameters are fixed with $k=10^{-14}$, $\dot{M}=10^{-8}\, M_{\odot} {\rm yr}^{-1}$, $\chi=10^{\circ}$ and $R=10^4\, {\rm km}$ at $t=0$. See~\cite{BM2017}.}%
 \label{fig3}
\end{figure}

If the magnetic field is dipolar, $\dot{\Omega} \propto \Omega^3$ for a fixed magnetic field. Generalizing it to $\dot{\Omega} = k\Omega^n$ with constant $k$ giving for the spin-powered phase, we obtain
\begin{eqnarray}
	\Omega &=& [\Omega_0^{1-n} - k(1-n)(t-t_0)]^{1/1-n}, \\
	{\rm and}\,\,\,B_s &=& \sqrt{\frac{5c^3 I k \Omega^{n-m}}{R^6 {\rm sin}^2 \chi}}.
\end{eqnarray}

Here, $\Omega_0$ is the initial angular velocity for the spin-powered phase (once accretion stops) at time $t=t_0$, and $\chi$ is the angle between the rotation
and magnetic axes. The value of $k$ is fixed such that $B_s$ can be constrained at $t=t_0$, which is known from the field evolution in the preceding accretion-powered phase. Here, $n=m=3$ corresponds to the dipole field configuration; therefore $m$ represents the deviation from the dipolar field, especially for $n=3$. Figure~\ref{fig3} shows the sample evolutions of angular velocity and magnetic field as functions of stellar mass. 
The initial angular velocity and particularly the magnetic field are chosen in such a way that they do not affect the stellar structure with~respect to when they are zero. In both cases, initially larger $\Omega$ with accretion drops significantly during the spin-powered phase, followed by a phase of its increasing trend. At~the end of the evolution, the~star can be left either as a super-Chandrasekhar WD and/or an SGR/AXP candidate with a higher spin frequency. Other initial conditions do not produce qualitatively different results.

\section{Simulating Magnetized and Rotating White Dwarfs and Their~Stability}
\label{Sec3}
In nature, WDs are expected to consist of mixed field geometry, including both toroidal and poloidal magnetic field components. However, here we consider toroidally dominated magnetic fields as they ensure the stability of these stars. 
Note, in fact, that the core of WDs may contain a stronger toroidal field,
which contributes to the stellar structure, mostly including mass, with~a weaker 
surface poloidal field. We simulate WDs and B-WDs based on the Einstein equation solver, \emph{XNS} code~\cite{Pili2014}.
It has been shown that toroidally dominated and purely toroidal fields not only make the star slightly prolate but also increases its equatorial radius~\cite{KS2004,FR2012,SM2015}. It is seen that the deformation at the core is more prominent than the outer region. Nevertheless, the rotation of a star makes it oblate, and hence, there is always a competition between these two opposing effects to decide whether the star will be an overall oblate or~prolate.

The combined effect of toroidal field and rotation (uniform and 
differential) was already explored in detail earlier \cite{Kalita2019}. 
It was shown that as the angular frequency is small, it does not affect the 
star considerably and results in a marginally prolate star due to magnetic 
effects. However, for high angular velocity, the low density region is affected more due to rotation than the high density central region, resulting in an 
oblate shaped star.
For both the explorations, the~magnetic to gravitational energies ratio (ME/GE) as well as kinetic to gravitational energies ratio (KE/GE) are chosen to be $\lesssim 0.1$ to maintain stable equilibrium~\cite{ChF1953,Komatsu1989,Braithwaite2009}. The authors found that for the central density $\rho_c \sim 2.2\times10^{10}\, {\rm g\ cm^{-3}}$, the magnetic field at the interior (close to the center) of the WD is $B_{max} \sim 2.7\times10^{14}\, {\rm G}$. Several WDs are observed with the surface magnetic field $\sim 10^9\, {\rm G}$
\cite{Heyl2000,Van2005,Brink2013}; hence, the central field might be much larger than $10^9\, {\rm G}$. In~fact, it has already been argued in the literature that the central field could be as large as $10^{14}\text{--}10^{15}\, {\rm G}$ \cite{FS2015,SS2017}.


For differentially rotating B-WDs~\cite{SM2015}, the~angular velocity profile is specified as~\cite{Ster2003,BDZ2011}
\begin{equation}
F(\Omega) = A^2 (\Omega_c - \Omega) = \frac{R^2 (\Omega - \omega)}{\alpha^2 - R^2 (\Omega- \omega)^2},
\end{equation}
which is implemented in the \emph{XNS} code used to simulate them,
where $A$ is a constant that indicates the extent of differential rotation, $R = \psi^2 r {\rm sin}\theta$, $\omega = -\beta^{\phi}$, $\Omega_c$ is the angular velocity at the center and $\omega$ is the angular velocity at radius $r$. 
Previous works (e.g. \cite{DM2015,SM2015,Kalita2019}) reported in detail the density isocontours 
of differentially rotating B-WDs for toroidal and poloidal fields. 
It was shown that `polar hollow' structure can form with differential rotation regardless of the specific geometry of the magnetic field.


\section{Mass-Limit and Luminosity Suppression in Non-Rotating White~Dwarfs}
\label{Sec4}
Apart from increasing the limiting mass of WDs, strong magnetic fields can also influence the thermal properties of the star, such as its luminosity, temperature gradient and cooling rate~\cite{MB2018,MB_euro2018,AG2020,MB2021,MB2022}. To~model such a WD, the~total pressure inside the star is modelled by including the contributions from the degenerate electron gas, ideal gas and magnetic pressures. The~interface is at the radius where the contributions from electron degenerate core and outer ideal gas pressures are the same. The~presence of strong fields gives rise to additional pressure $P_B = B^2/8\pi$ and density $\rho_B = B^2/8\pi c^2$ inside the magnetized WDs~\cite{Sinha2013} when $c$ is the speed of~light.

For an approximately spherical B-WD, the~model equations for magnetostatic equilibrium, photon diffusion and mass conservation can be written within a Newtonian framework as
\begin{eqnarray} 
&\frac{{\rm d}}{{\rm d}r}(P_{\rm deg}+P_{\rm ig}+P_{B}) = -\frac{Gm(r)}{r^2}(\rho + \rho_B), \nonumber \\
&\frac{{\rm d}T}{{\rm d}r} = -{\rm max}\left[\frac{3}{4ac} \frac{\kappa \rho}{T^3} \frac{L_r}{4\pi r^2}, \left(1 - \frac{1}{\gamma}\right)\frac{T}{P} \frac{{\rm d}P}{{\rm d}r}\right], \vspace{0.1in} \\
&\frac{{\rm d}m}{{\rm d}r} = 4\pi r^2 (\rho + \rho_B), \nonumber
\end{eqnarray}
where we have ignored the magnetic tension term for radially varying magnetic field strength. In~these equations, $P_{\rm deg}$ and $P_{\rm ig} = \rho kT/\mu m_p$ are the electron degeneracy pressure and the ideal gas pressure, $\rho$ is the matter density, $T$ is the temperature, $m(r)$ is the mass enclosed within radius $r$, $\kappa$ is the radiative opacity, $L_r$ is the luminosity at radius $r$, and~$\gamma$ is the adiabatic index of the~gas. 

The opacity in the surface layers of non-magnetised WD is approximated with Kramers' formula, $\kappa = \kappa_0 \rho T^{-3.5}$, where $\kappa_0 = 4.34\times10^{24}Z(1+X)\ {\rm cm^2\ g^{-1}}$, and~$X$ and $Z$ are the mass fractions of hydrogen and heavy elements (other than hydrogen and helium) in the stellar interior, respectively. Assuming helium WDs for our study here, we set the helium mass fraction to $Y = 0.9$ and $Z = 0.1$. The~radiative opacity in the surface layers is mainly due to the bound-free and free-free transitions of electrons~\cite{ST1983}. 
For the radial variation of the field magnitude within the B-WD, we adopt a profile used extensively to model magnetized NSs and B-WDs~\cite{Band1997,DM2014},
given by
\begin{equation}
B\left(\frac{\rho}{\rho_0}\right) = B_{\rm s} + B_0 \left[1 - {\rm exp}\left(-\eta \left(\frac{\rho}{\rho_0}\right)^{\gamma}\right)\right],
\label{Bprof}
\end{equation}
where $B_s$ is the surface magnetic field, $B_0$ is a fiducial magnetic field, and $\eta$ and $\gamma$ are dimensionless parameters along with $\rho_0=10^9\, {\rm g/cm^3}$, which determine how the magnetic field decays from the degenerate core to the surface. We set $\eta = 0.8$ and $\gamma = 0.9$ for all~calculations, unless stated otherwise.

Radial luminosity is assumed to be constant, $L_r = L$, as~there is no hydrogen burning or other nuclear fusion reactions that take place within the WD core. The~differential equations are solved by providing the matter density at the surface, total WD mass and surface temperature as the boundary conditions. For~strong magnetic fields, the~variation of radiative opacity with $B$ can be modelled as $\kappa = \kappa_B \approx 5.5\times10^{31}\rho T^{-1.5} B^{-2} {\rm cm^{2}\ g^{-1}}$~\cite{PY2001,VP2001}. The~field-dependent Potekhin's opacity is used instead of Kramers' opacity if $B/10^{12}\ {\rm G} \geq T/10^6\ {\rm K}$, which is relevant for the strong fields that we~consider. 

\begin{figure}[H]
\begin{tabular}{cc}
\includegraphics[width=2.6in]{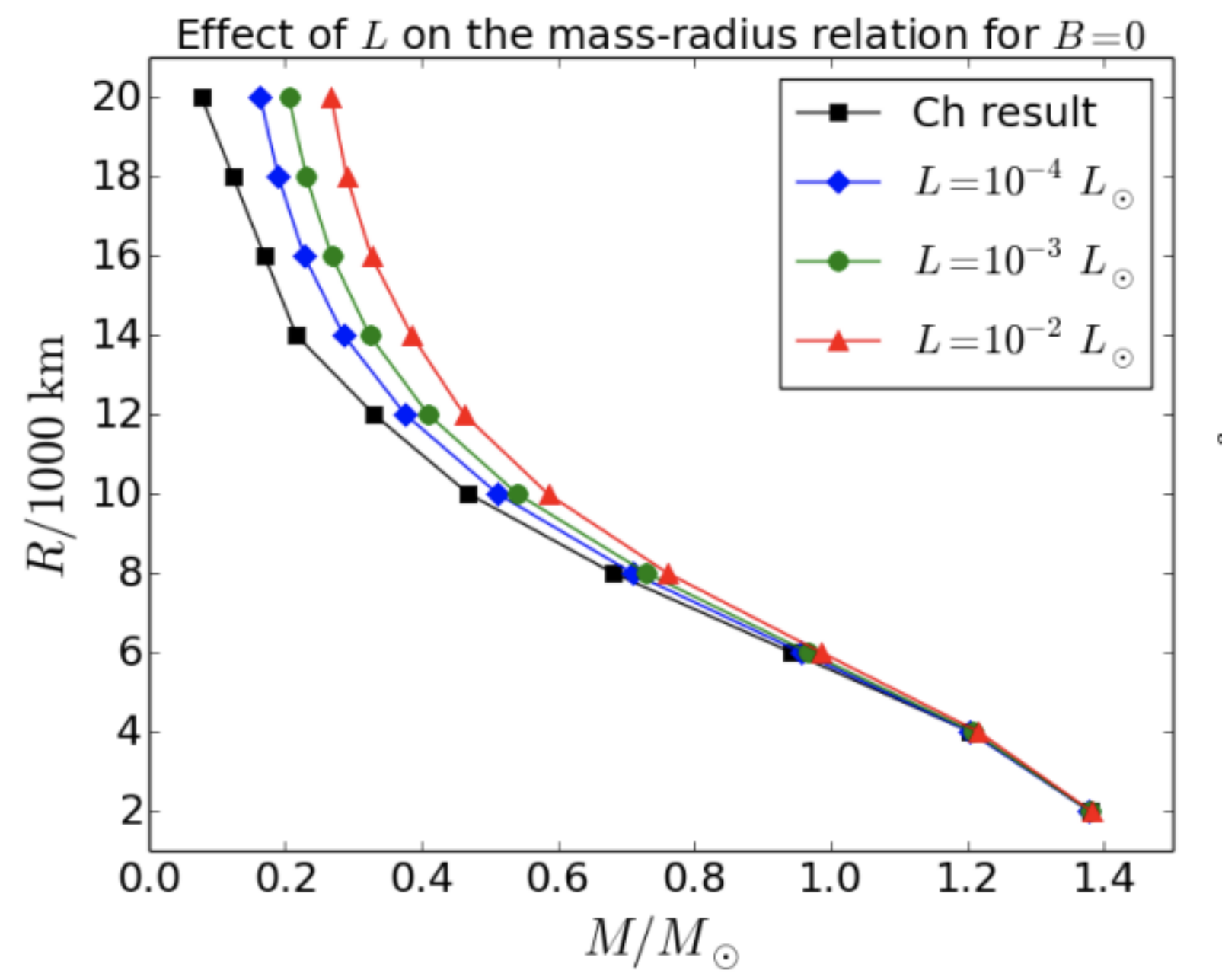}
&\includegraphics[width=2.6in]{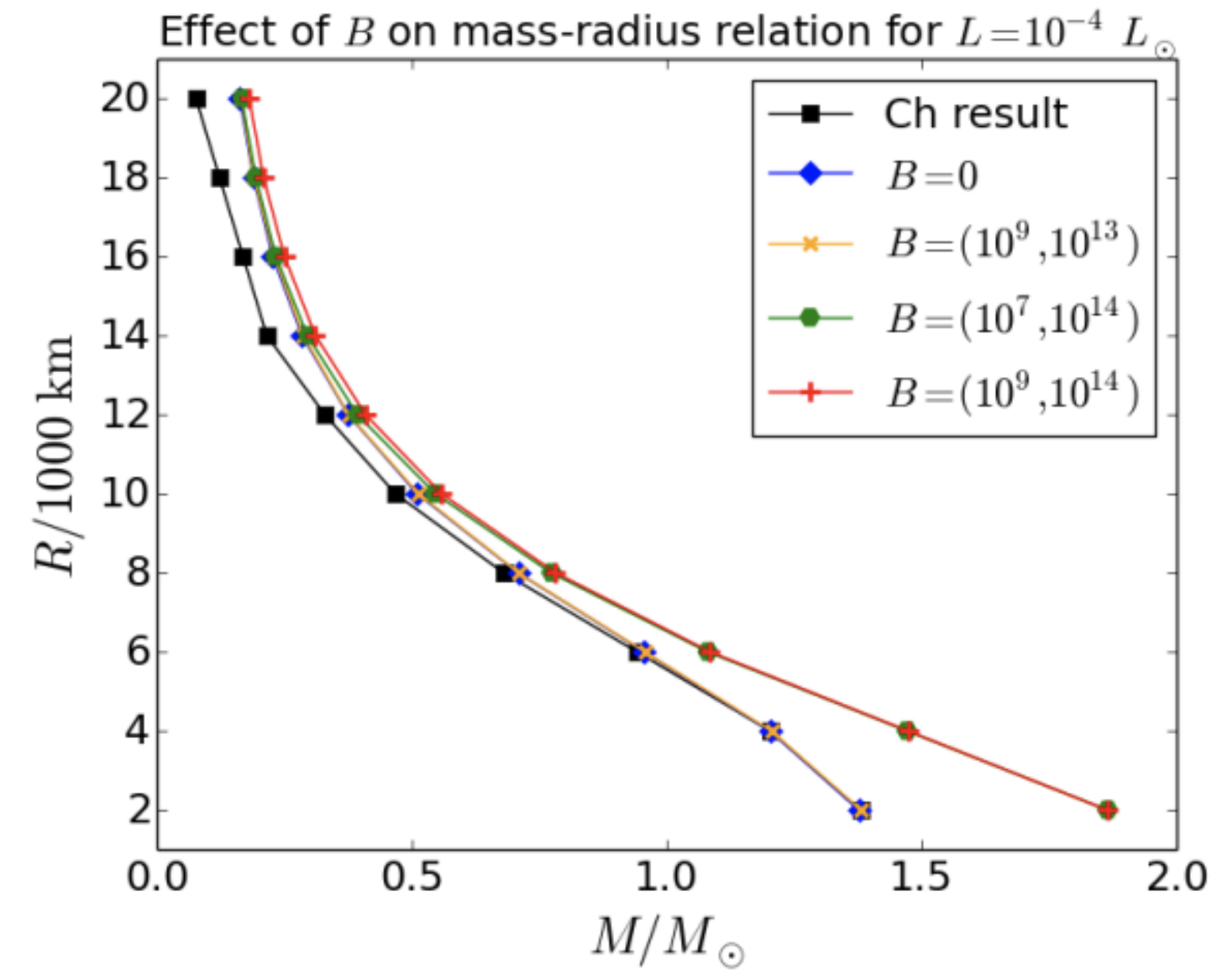}\\
\end{tabular}
  \caption{\emph{Left panel:} The effect of $L$ on the mass--radius relation of non-magnetised WDs is shown for $L=10^{-4}\ L_{\odot}$ (blue diamonds), $L=10^{-3}\ L_{\odot}$ (green circles) and $L=10^{-2}\ L_{\odot}$ (red triangles), along with the Chandrasekhar result (black squares). \emph{Right panel:} The effect of field strength on the mass--radius relation of B-WDs is shown for 
	$B=(B_s,B_0)=(0,0)$ (blue diamonds), $B=(10^9,10^{13})\ {\rm G}$ (orange crosses), $B=(10^7,10^{14})\ {\rm G}$ (green circles) and $B=(10^9,10^{14})\ {\rm G}$ (red pluses), along with the Chandrasekhar result (black squares) for $L=10^{-4}\ L_{\odot}$. See~\cite{MB2022}.}%
\label{fig6}
\end{figure}

The left panel of Figure~\ref{fig6} shows the effect of luminosity on the mass-radius relation for non-magnetized WDs compared to Chandrasekhar's results~\cite{Ch1935}. Although~the increase in $L$ leads to progressively higher masses for larger WDs, Chandrasekhar mass limit is retained irrespective of the luminosity. The~right panel of Figure~\ref{fig6} shows the effect of magnetic field on the mass-radius relation for B-WDs with $L=10^{-4}\, L_{\odot}$ and compares them with the non-magnetic Chandrasekhar results. It can be seen that the magnetic field affects the mass-radius relation in a manner analogous to increasing $L$ by~shifting the curve towards higher masses for these WDs having larger radii. The~mass-radius curves for $B_0 \lesssim 10^{13}\, {\rm G}$ practically overlap with each other in the smaller radius regime and retain the Chandrasekhar mass limit. However, for~strong central fields with $B_0 \sim 10^{14}\, {\rm G}$, super-Chandrasekhar WDs are obtained, with masses as high as $\sim$$1.9\, M_{\odot}$.

To ensure structural stability of B-WD, an~increase in magnetic energy density has to be compensated by a corresponding decrease in the thermal energy and hence the luminosity. This effect is especially prominent for B-WDs with larger radii where the magnetic, thermal and gravitational energies are comparable with each other. We find that in the presence of stronger field, a~slight decrease in the luminosity for $R \gtrsim 12000\, {\rm km}$ WDs leads to masses that are similar to their non-magnetic counterparts. However, the~smaller radius B-WDs require a substantial drop in their luminosity (well outside the observable range) and still do not really achieve masses that are similar to the non-magnetized WDs. 
As a result, for~stars with $2000 \leq R/{\rm km} \leq$~10,000, even if the luminosity decreases substantially with $10^{-16} \leq L/L \leq 10^{-12}$, the~resulting mass of the B-WD remains larger than its non-magnetic counterpart. This leads to an extended branch in the mass-radius~relation.

\section{Effect of Magnetic Field Dissipation and Cooling~Evolution}
\label{Sec5}
Magnetic fields inside WD undergo decay by Ohmic dissipation and Hall drift processes with timescales given by~\cite{HK1998,Cumming2002}, respectively,
\begin{eqnarray}
t_{\rm Ohm} = (7\times10^{10}\ {\rm yr})\, \rho_{c,6}^{1/3} R_{4}^{1/2} (\rho_{\rm avg}/\rho_{\rm c}),\\
t_{\rm Hall} = (5\times10^{10}\, {\rm yr})\ l_8^2 B_{0,14}^{-1} T_{\rm c,7}^{2} \rho_{\rm c,10},
\label{tOhm_tHall} 
\end{eqnarray}
where $\rho_{\rm c,n} = \rho_c/10^n\, {\rm g\, cm^{-3}}$, $R_4 = R/10^4\, {\rm km}$, $T_{c,7}=T_c/10^7\, {\rm K}$, $B_{0,14}=B_0/10^{14}\, {\rm G}$ and $l = l_8 \times 10^8\, {\rm cm}$ is the characteristic length scale of the flux loops through the outer core of WD. Ohmic decay is the dominant field dissipation process for $B \lesssim 10^{12}\ {\rm G}$, while for $10^{12} \leq B/{\rm G} \leq 10^{14}$, the decay occurs via Hall drift, and for $B \gtrsim 10^{14}\ {\rm G}$, the~principal decay mechanism is likely to be ambipolar diffusion~\cite{HK1998}. 

The field decay can be solved using
\begin{equation}
\frac{{\rm d}B}{{\rm d}t} = -B\left(\frac{1}{t_{\rm Ohm}} + \frac{1}{t_{\rm Amb}} + \frac{1}{t_{\rm Hall}}\right),
\end{equation}
where $t_{\rm Amb}$ denotes the ambipolar diffusion time scale. We consider two separate cases: (a) when only Ohmic dissipation occurs for both the surface and central magnetic fields, and (b) while $B_s$ continues to evolve over $t_{\rm Ohm}$, Hall drift determines the $B_0$ evolution until the central field drops to about $10^{12}\, {\rm G}$,
below which Ohmic dissipation sets~in.

The thermal energy is radiated away gradually over time in the observed luminosity from the surface layers as the star evolves. Because~most of the electrons occupy the lowest energy states in a degenerate gas, the~thermal energy of ions is the only significant energy source that can be radiated. The~rate at which thermal energy of ions is transported to surface and radiated depends on the specific heat, given by
\begin{equation}
L = -\frac{d}{dt}\int c_{v} dT = (2\times10^6\ {\rm erg/s)}\, \frac{Am_{\mu}}{M_{\odot}}\left(\frac{T}{K}\right)^{7/2},
\end{equation}
where $c_v \approx 3k_B/2$ is the specific heat at constant volume. Given an initial $L$ and temperature $T_0$ at time $t_0$, final temperature after cooling is given by $(T/{\rm K})^{-5/2} - (T_0/{\rm K})^{-5/2} = 2.406\times10^{-34}\, \tau/{\rm s}$, where $\tau = t-t_0$ is the WD age.  It is important to note that for simplicity in calculations, we have assumed self-similarity of the cooling process over the entire~evolution.

Table~\ref{Table1} lists the luminosities and masses for WDs with radii $2000 \lesssim R/{\rm km} \lesssim$~20,000 and initial $B=(10^9,10^{14})\, {\rm G}$ at~time $t=0$ and $10\, {\rm Gyr}$. The~fraction of time when the Hall drift dominates, i.e.,~$t_{\rm Hall}/\tau$, falls significantly with increasing stellar radius. Therefore, the~magnetic field decays considerably more because the faster Ohmic dissipation process turns out to be critical for much of the cooling evolution. The~mass-radius relations merge for $R \gtrsim 6000\, {\rm km}$ WDs. The~inferred luminosities are also much less suppressed for the intermediate radius WDs with $6000 \lesssim R/{\rm km} \lesssim 12000$. However, for~low radius B-WDs,
field decay affects mass and luminosity significantly. As~it is the high 
magnetic pressure which helps to hold more mass, the~field decay significantly
sheds off mass in its new equilibrium if the radius is fixed. The~limiting 
mass for small radius B-WDs with a fixed $R \approx 2000\, {\rm km}$ drops to about 
$1.5\ M_{\odot}$ due to field decay compared to $1.9\ M_{\odot}$ without field evolution. 
The majority of these small radius B-WDs still remain practically hidden throughout their cooling evolution because of their strong fields and correspondingly low luminosity. 

\begin{table}[H]
\caption{The effect of magnetic field on luminosity when the initial field is fixed at $B=(10^9,10^{14})\ {\rm G}$ for all the radii. The topmost entry for each radius is the initial time, whereas the bottom two entries list corresponding parameters for $t=\tau=10\ {\rm Gyr}$. We evaluate fields assuming Ohmic dissipation is the dominant process for the top entries of $\tau=10\ {\rm Gyr}$, and~for the bottom entries, we assume Hall drift is the primary process until the field parameter $B_0$ decays to $\sim$$10^{12}\ {\rm G}$, below~which Ohmic dissipation dominates.
}
\label{Table1}
\bgroup
\def\arraystretch{1.2}
\begin{tabular}{c c c c c c c c }
\hline
\hline
\centering
\textbf{\emph{R}/1000 km} & \boldmath{$t/{\rm Gyr}$} & \boldmath{$t_{\rm Hall}/\tau$ }
& \boldmath{$B_{\rm s}/{\rm G}$} & \boldmath{$B_0/{\rm G}$} & \boldmath{$L/L_{\odot}$ }& \boldmath{$M_{B=0}/M_{\odot}$ }& \boldmath{$M/M_{\odot}$} \\ \midrule 
2.0 & 0 & & $10^9$ & $10^{14}$ & $10^{-16}$ & 1.378 & 1.865 \\ \noalign{\smallskip} \cmidrule{2-8} \noalign{\smallskip}
    & 10 & 0 & $4.58\times10^8$ & $4.58\times10^{13}$ & $10^{-16}$ & 1.377 & 1.478 \\
    & & 1 & & $5.83\times10^{13}$ & $10^{-16}$ & & 1.542 \\ \midrule
8.0 & 0 & & $10^9$ & $10^{14}$ & $10^{-12}$ & 0.709 & 0.762 \\ \noalign{\smallskip} \cmidrule{2-8} \noalign{\smallskip}
    & 10 & 0 & $9.86\times10^7$ & $9.86\times10^{12}$ & $2\times10^{-6}$ & 0.699 & 0.699 \\
    & & $2.28\times10^{-2}$ & & $1.04\times10^{11}$ & $7\times10^{-6}$ & & 0.699 \\ \midrule
14.0 & 0 & & $10^9$ & $10^{14}$ & $2\times10^{-6}$ & 0.286 & 0.286 \\ \noalign{\smallskip} \cmidrule{2-8} \noalign{\smallskip}
    & 10 & 0 & $3.06\times10^7$ & $3.06\times10^{12}$ & $10^{-5}$ & 0.262 & 0.262 \\ 
    & & $2.22\times10^{-3}$ & & $3.08\times10^{10}$ & $10^{-5}$ & & 0.262 \\ \midrule
20.0 & 0 & & $10^9$ & $10^{14}$ & $7\times10^{-6}$ & 0.164 & 0.164 \\ \noalign{\smallskip} \cmidrule{2-8} \noalign{\smallskip}
    & 10 & 0 & $2.97\times10^7$ & $2.97\times10^{12}$ & $10^{-5}$ & 0.138 & 0.138 \\ 
    & & $3.70\times10^{-4}$ & & $2.98\times10^{10}$ & $10^{-5}$ & & 0.138 \\ 
			\bottomrule
\end{tabular}
\egroup
\end{table}

\begin{figure}[H]
\begin{tabular}{cc}
\includegraphics[width=2.6in]{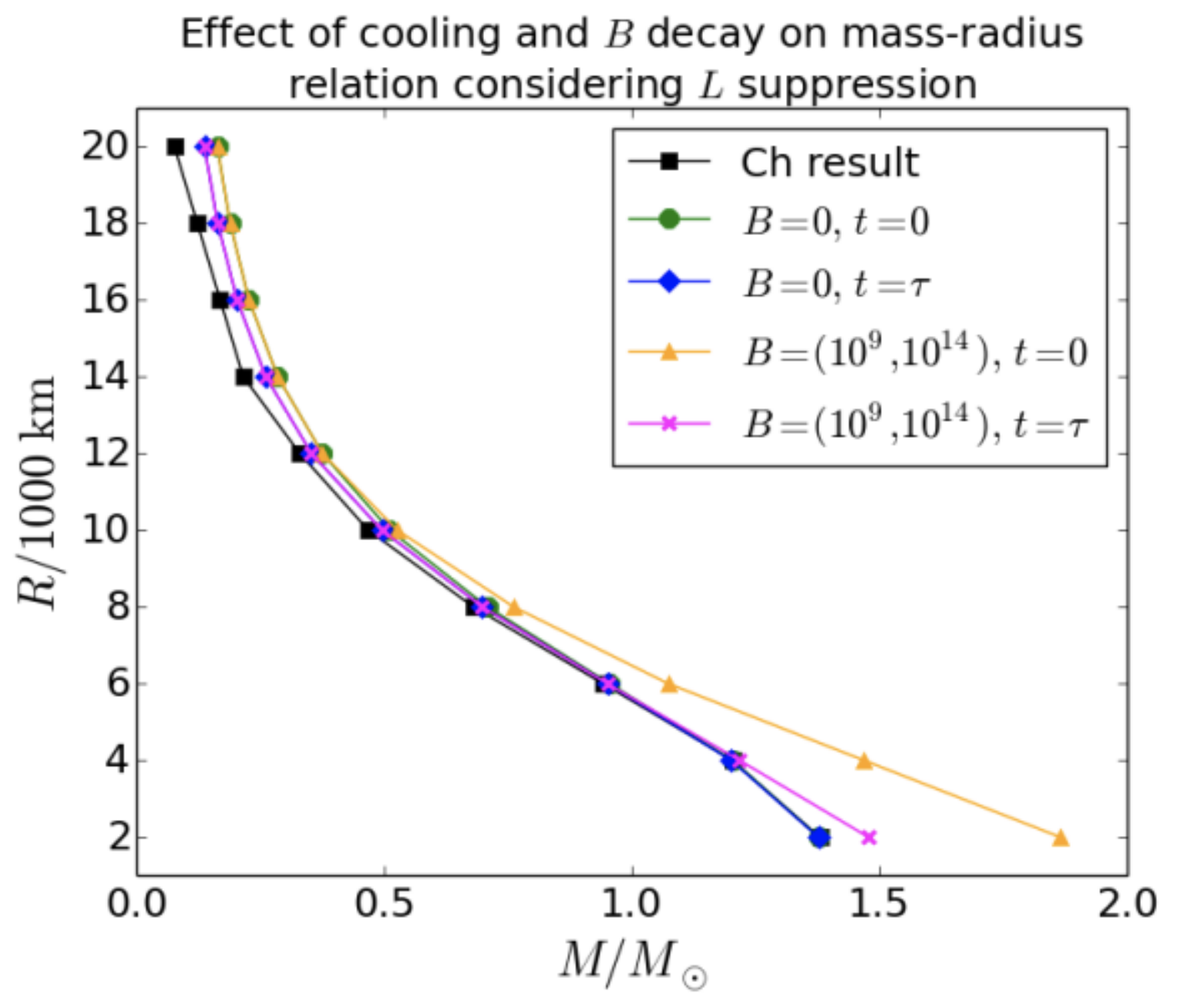}
&\includegraphics[width=2.6in]{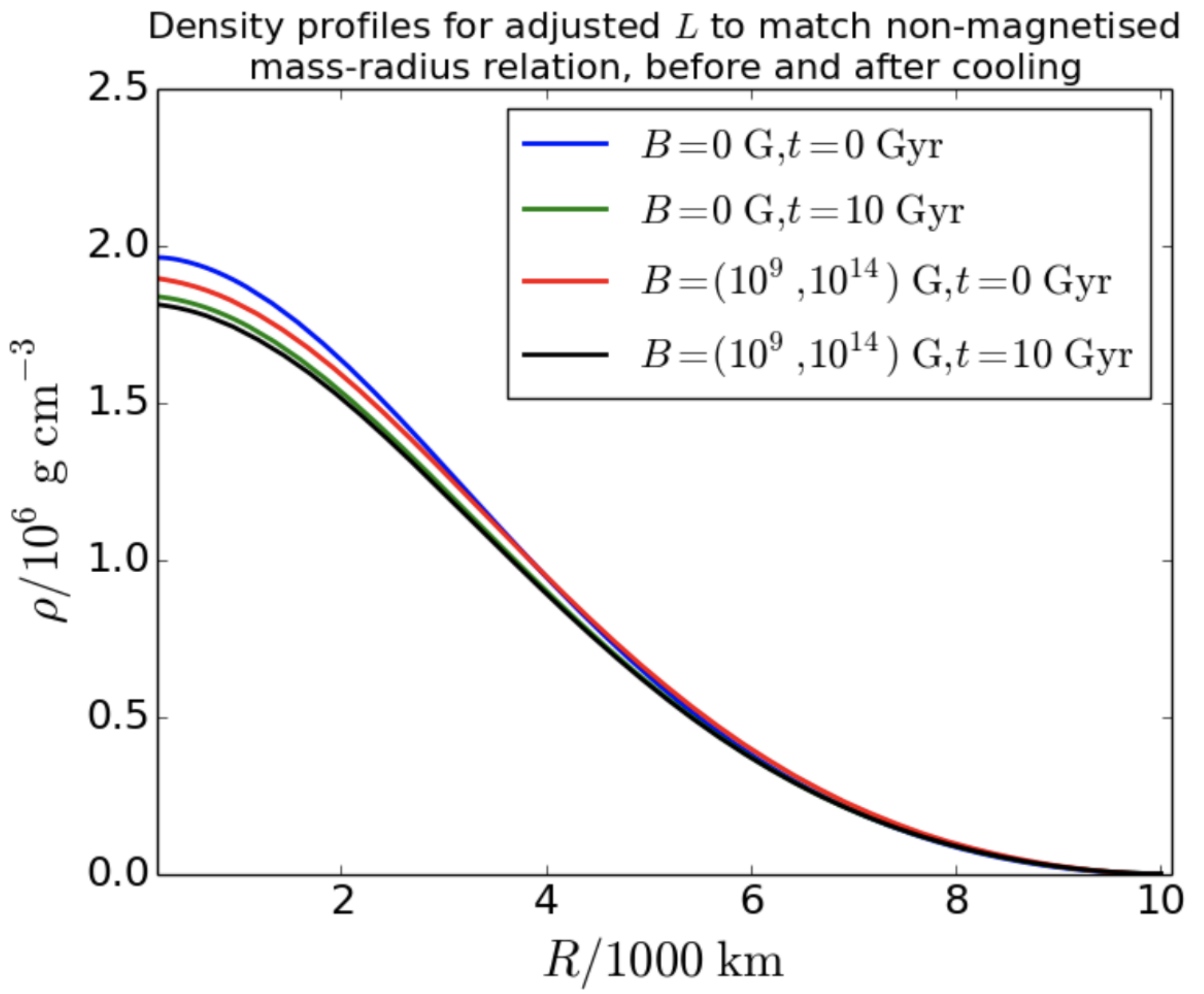}\\ 
\end{tabular}
	\caption{\emph{Left panel:} Effect of magnetic field on $L$ set to match the non-magnetised mass-radius relation. Results are shown for $B=(0,0)$ at $t=0$ (green circles), $B=(0,0)$ at $t=10\ {\rm Gyr}$ (blue diamonds), $B=(10^9,10^{14})\, {\rm G}$ at $t=0$ (orange triangles) and $B=(10^9,10^{14})\, {\rm G}$ at $t=10\ {\rm Gyr}$ (magenta crosses). \emph{Right panel:} The matter density profiles for the same cases are shown for \emph{R} = 10,000 km. See~\cite{MB2022}.}%
\label{fig7}
\end{figure}

The left panel of Figure~\ref{fig7} shows the effect of B-WD evolution on their mass-radius relations, including both magnetic field decay and thermal cooling effects but neglecting neutrino cooling. 
The luminosities are varied with field strength such that the B-WD masses can match those obtained for the non-magnetized WDs. For~$B=(0,0)\, {\rm G}$, the~mass-radius relation is shifted more towards the Chandrasekhar result as a result of cooling, and the mass limit remains unaltered. However, for~$B=(10^9,10^{14})\, {\rm G}$, although~the maximum mass $\sim$$1.9\, M_{\odot}$ at a small radius turns out to be much larger than the Chandrasekhar limit, we find that it is lowered considerably to $\sim$$1.5\, M_{\odot}$ primarily as a result of magnetic field decay and also thermal cooling over $t=10\, {\rm Gyr}$. The~right panel of Figure~\ref{fig7} shows the radial variation of matter density for the same cases. Matter density at the core is slightly suppressed in the presence of strong field and also as a result of the evolution. As~the total stellar energy is conserved, an~increase in the magnetic energy has to be compensated by a similar decrease in gravitational energy, and, hence, the central density. Once the field decays and luminosity drops due to cooling, the~central density adjusts itself to be slightly lower to balance the loss of magnetic and thermal energies with~time.

We also use the STARS stellar evolution code to qualitatively investigate the B-WD mass-radius relationship at different field strengths, with~the objective of numerically validating our semi-analytical models. We find that the numerical results are in good agreement with our analytical formalism, and the magnitude of $B_0$ dictates the shape of the mass-radius curve. In~validation of our analytical approach, we have found that the limiting mass $\sim$$1.8703 M_{\odot}$ obtained with the STARS numerical models is in very good agreement with $M \approx 1.87 M_{\odot}$, which is inferred from the semi-analytical calculations for B-WDs with strong fields $B=(10^{\text{6--9}},10^{14})\, {\rm G}$ \cite{MB2022} for a given magnetic field profile. 

We argue that the young super-Chandrasekhar B-WDs only sustain their large masses up to $\sim$$10^{5}\text{--}10^6\, {\rm yr}$ since their formation, and this essentially explains their apparent scarcity even without the difficulty of detection owing to their suppressed luminosities. We plan to explore this
issue in detail in the future, particularly the timescale of their formation
considering simultaneously the growth (e.g., by accretion~\cite{BM2017}) and
decay of fields. 
It is important to note that if matter accretes at a higher 
rate such that the total mass accreted exceeds $\sim$0.1--0.2$M_\odot$ before
the field diffuses, 
the field may be restructured~\cite{zwf}, leading to a reduced polar field
independent of the initial field. We also plan to rigorously explore the 
fate of B-WDs once the fields decay, if~they actually collapse, in~place of 
keeping the radius~fixed.

\section{Mass Limit Estimate Including Quantum~Effects}
\label{Sec6}

\textls[-15]{Several magnetized WDs have been discovered with surface fields as high as $\sim$$10^5\text{--}10^9\, {\rm G}$}. It is likely that stronger fields ($\sim$$10^{12}\text{--}10^{14}\, {\rm G}$) exist at their interiors. Energy states of a free electron in a uniform strong magnetic field are quantized into Landau orbitals, which defines the motion of the electron in a plane perpendicular to the field. The maximum number of Landau levels occupied by cold electrons in a magnetic field is given by $\nu_m = \frac{(E_{Fmax}/m_e c^2)^2 - 1}{2 B_D}$, where $m_e$ is the rest mass of the electron, $E_{Fmax}$ is the maximum Fermi energy of the system and $B_D = B/B_c$, where $B_c = 4.414\times10^{13}\, {\rm G}$, is the critical field strength. High magnetic field strength therefore modifies the equation of state (EoS) of degenerate matter by causing Landau quantization of electrons~\cite{LS1991}. The larger the magnetic field, the smaller is the number of Landau levels occupied~\cite{DM2013}. This results in a significant modification of the mass-radius relation of the underlying WD, particularly for $B_D\gtrsim 100$. 

To obtain the revised mass-radius relation for a strongly magnetized WD, the~modified EoS has to be combined with the condition of magnetostatic equilibrium. If~the B-WD is approximated to be spherical, then its mass is obtained from
\begin{equation}
\frac{1}{\rho+\rho_B}\frac{d}{dr}\left(P+\frac{B^2}{8\pi}\right) = -\frac{GM}{r^2} + \left[\frac{\vec{B}\ . \nabla \vec{B}}{4\pi (\rho+\rho_B)}\right]_{r}, \ \frac{dM}{dr} = 4\pi r^2 (\rho + \rho_B).
\end{equation}

If the field is uniform or highly fluctuating, the~magnetic terms can be neglected in the above equations, and~following Lane--Emden formalism~\cite{ARC2010}, the~scalings of mass and radius with central density are obtained as
\begin{equation}
M \propto K_m^{3/2}\rho_c^{(3-n)/2n}, \ R \propto K_m^{1/2} \rho_c^{(1-n)/2n}, \ K_m = K\rho_c^{-2/3}.
\end{equation}

\begin{figure}[H]
\begin{tabular}{cc}
\includegraphics[width=2.45in]{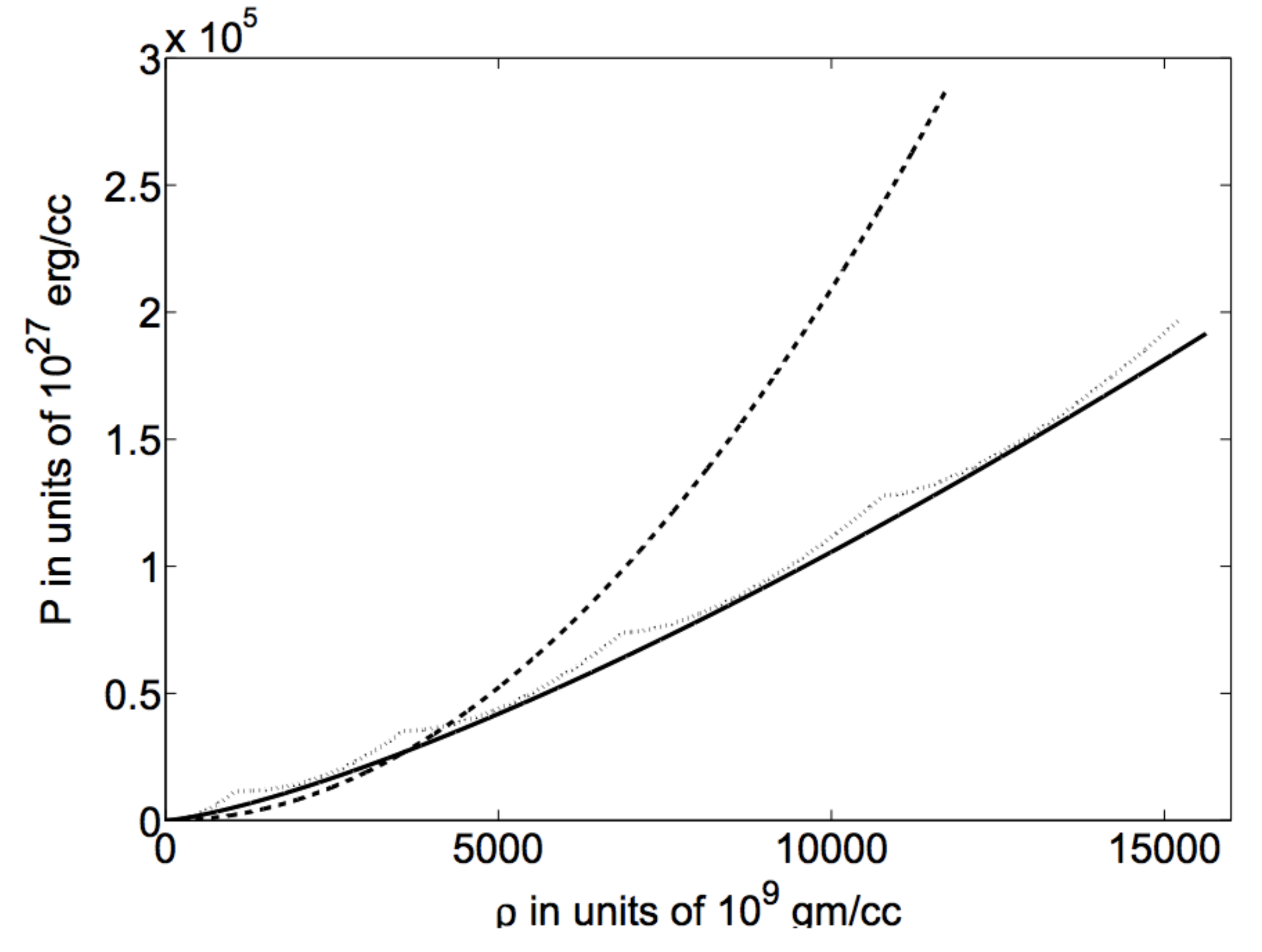}
&\includegraphics[width=2.69in]{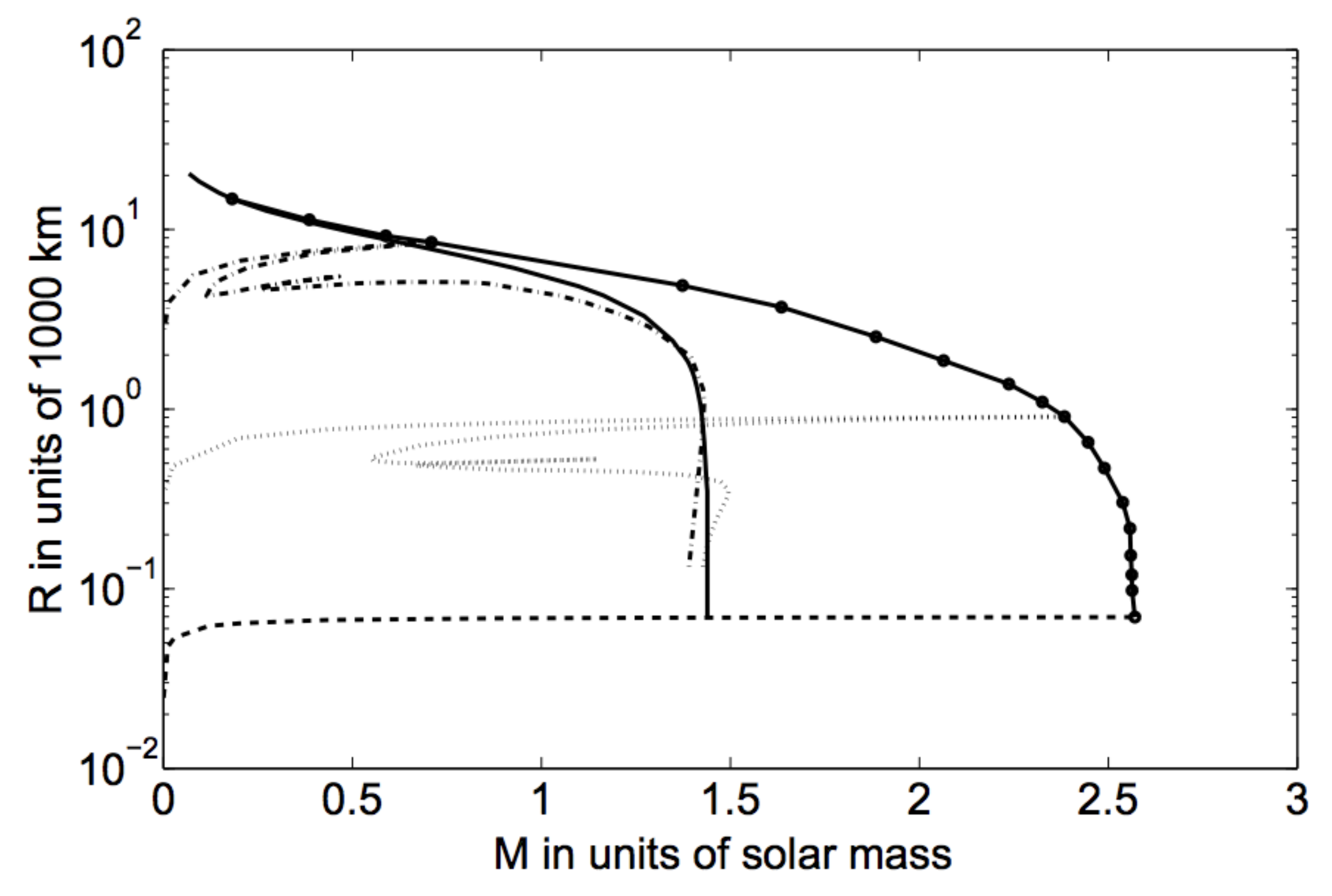}\\
\end{tabular}
	\caption{\emph{Left panel:} EoS for different cases: solid line represents Chandrasekhar's EoS, and the~dotted and dashed lines represent the 5-level (corresponding to very strong $B$) and 1-level (corresponding to ultimate EoS for extreme $B$) systems of Landau quantization, respectively. \emph{Right panel:} Mass-radius relations: pure solid line represents Chandrasekhar's result, and the one marked with filled circles represents the evolutionary track of WD with an increase in $B$. The~dot-dashed, dotted and dashed lines represent WDs with 50124-level, 200-level and 1-level systems of Landau quantization, respectively (corresponding to increasing $B$). We set $E_{Fmax}=200 m_e c^2$ for both panels. See~\cite{DM2013,DM2013b}.}%
\label{fig8}
\end{figure}

Here, $n=1$ ($\Gamma=2$), which is the case for a high $B_D$ as shown in 
Figure~\ref{fig8}, corresponds to central density independent mass, unlike Chandrasekhar's case when $K_m$ is independent of the field strength, and limiting mass corresponds to $n=3$. 

Substituting the proportionality constants appropriately, the~limiting mass is obtained~as
\begin{equation}
M_{lim} = \left(\frac{hc}{2G}\right)^{3/2} \frac{1}{(\mu_e m_H)^2} \approx \frac{10.312}{\mu_e^2}M_{\odot},
\end{equation}
when the limiting radius $R_{lim} \rightarrow 0$. For~$\mu_e=2$, which is the case for a carbon-oxygen WD, we obtain $M_{lim} = 2.58 M_{\odot}$. For~a finite but high-density and magnetic field, e.g.,~$\rho_c = 2\times10^{10}\, {\rm g\ cm^{-3}}$ and $B=8.8\times10^{15}\, {\rm G}$ when $E_{Fmax}=20m_e c^2$, $M_{lim}=2.44 M_{\odot}$ and $R \approx 650\, {\rm km}$. It should be noted that these $\rho_c$ and $B$ are similar or below their respective upper limits set by the instabilities of pycnonuclear fusion, inverse-$\beta$ decay and general relativistic effects~\cite{DM2014}. Note, however, that pycnonuclear reaction rates are quite
uncertain and not well constrained. Moreover,~slightly lower $\rho_c$ 
and $B$ still would lead to B-WD mass well above the Chandrasekhar~mass-limit.

The left panel of Figure~\ref{fig8} shows how the EoS of electron degenerate matter is modified as a result of magnetic field. With~increasing field, the~inward gravitational force is balanced by the outward force due to modified matter 
pressure, and a quasi-equilibrium state is attained. Consequently, with~
decreasing stellar radius due to a higher gravitational field, a~very high magnetic field is generated, which prevents the WD from collapsing, thus making its 
equilibrium configuration more massive. Subsequently, with~the continuation of accretion the WD approaches the new mass limit $M_{lim} \sim 2.58M_{\odot}$. 
This new limiting mass WD plausibly sparks off a violent thermonuclear reaction with further accretion, as is the case with the idea of the $1.4M_\odot$ limit of nonmagnetic WDs, 
thus exploding it and giving rise to an over-luminous type Ia~supernova. 

The evolution of the mass-radius relationship with the evolution of magnetic field 
for a super-Chandrasekhar WD of maximum possible mass is shown in the right panel of Figure~\ref{fig8}, along with a few typical mass-radius relations for different fixed magnetic field strengths describing possible stars in intermediate equilibrium states. The~ultimate WD, corresponding to the maximum mass $M_{lim} \sim 2.58M_{\odot}$, lies on the mass-radius relation for a one-Landau-level system, but~the intermediate WDs having weaker magnetic fields correspond to multilevel systems. The~said one-Landau-level system corresponds to the hypothetical high central magnetic field $8.8\times10^{17}\, {\rm G}$. This is in the spirit of a Chandrasekhar mass-limit for a non-magnetic WD, which also
corresponds to the hypothetical high (even infinity) central density. However, the~mass for
$B_D\sim 30-100$ still appears to be significantly super-Chandrasekhar. The~intermediate systems of 200-level and 50124-level systems correspond to central magnetic fields of $4.4\times10^{15}\, {\rm G}$ and $1.7\times10^{13}\, {\rm G}$, respectively.

\section{Detectability of Gravitational Waves from Magnetised White~Dwarfs}
\label{Sec7}
\begin{figure}[H]%
\begin{tabular}{cc}
\hspace{-30pt}\includegraphics[width=2.2in]{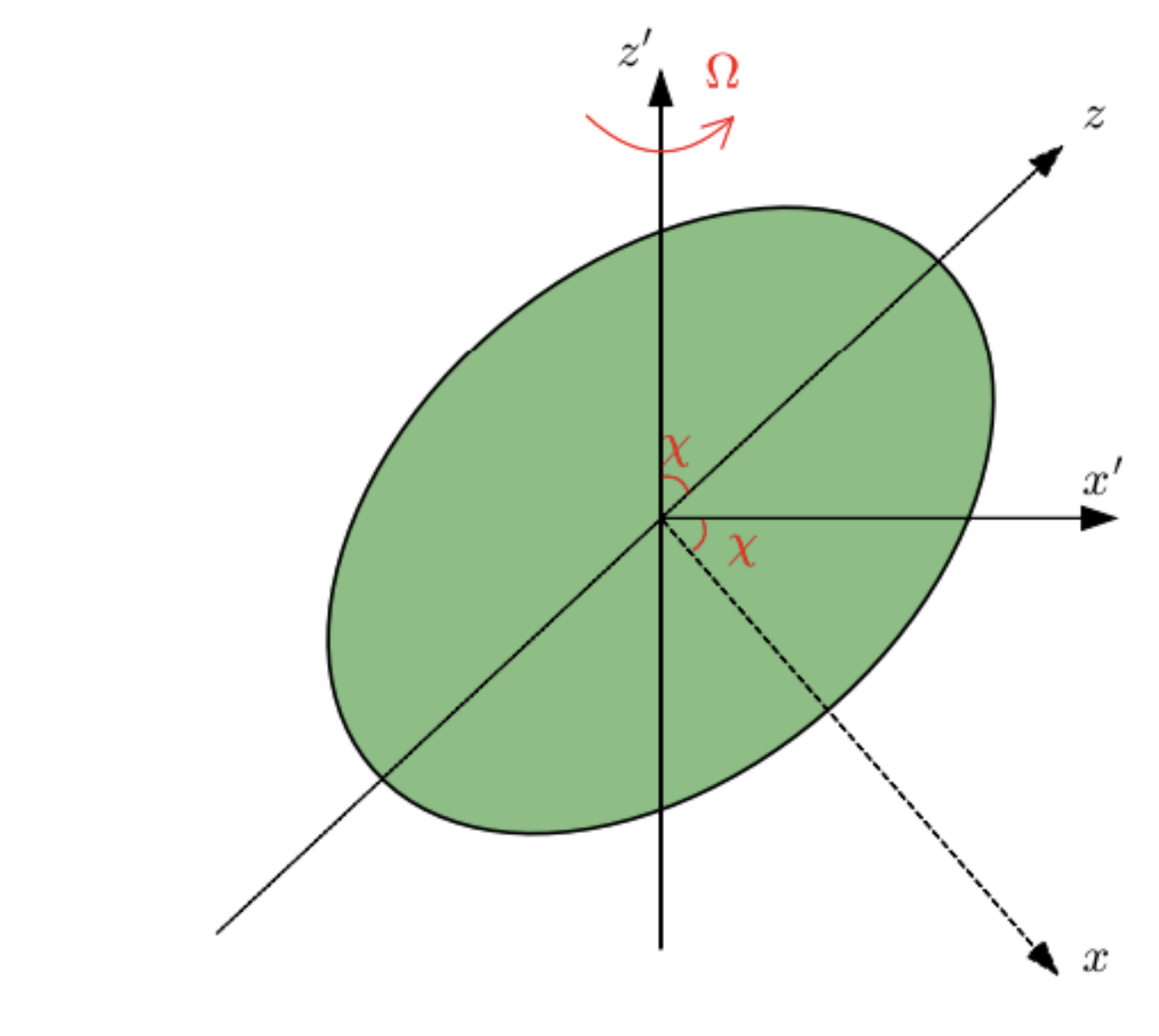}
&\includegraphics[width=2.8in]{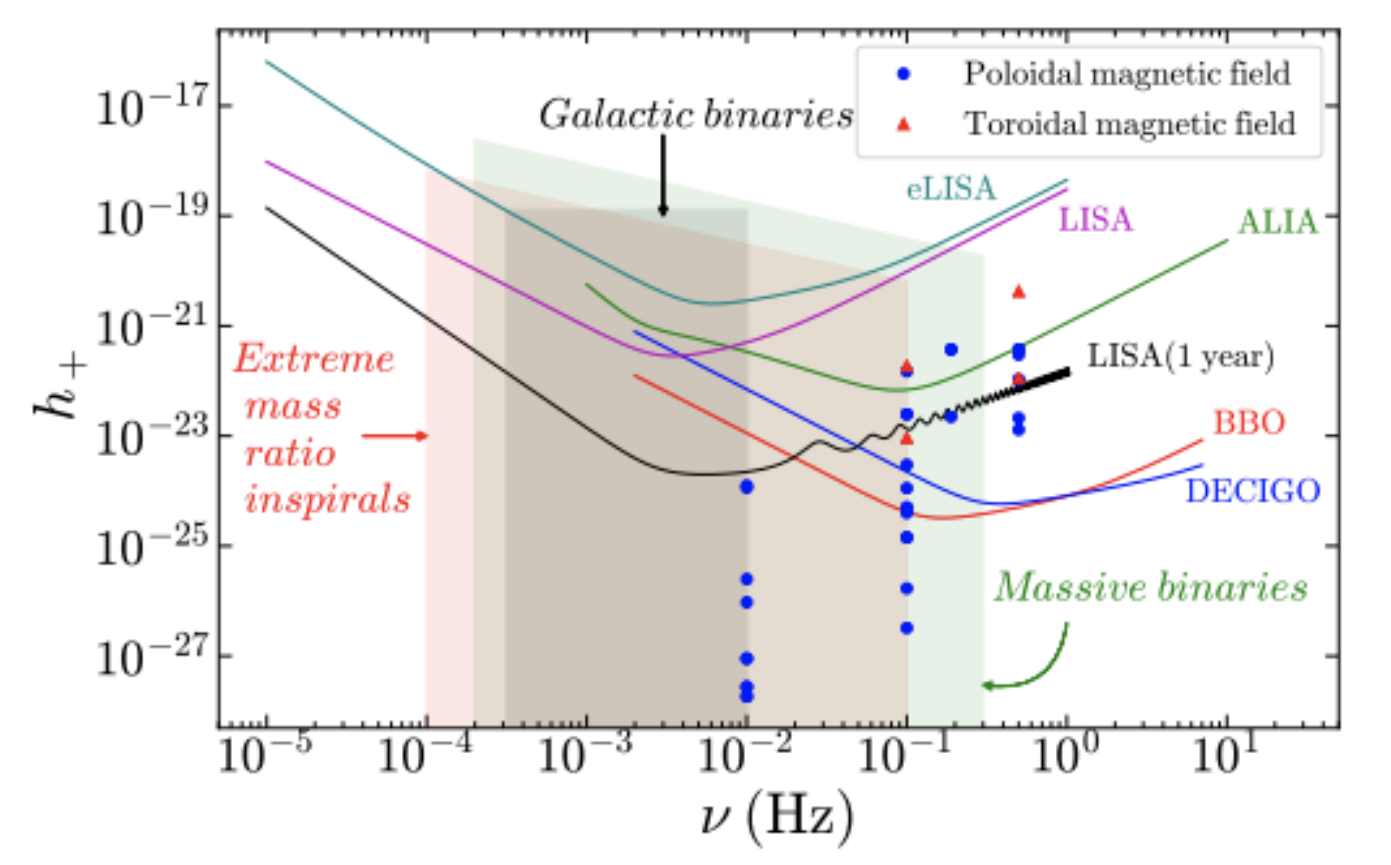}\\
\end{tabular}
	\caption{\emph{Left panel:} Schematic diagram of a B-WD with $z^{\prime}$ being the rotational axis and $z$ as the magnetic axis. \emph{Right panel:} The dimensionless GW amplitudes for WDs are shown as functions of frequency, along with the sensitivity curves of various detectors. Optimum $i$ is chosen for $\chi$ at $t=0$. See~\cite{Kalita2020}.}%
\label{fig9}
\end{figure}

One question that remains to be answered is how can B-WDs be detected directly.
Continuous gravitational waves can be among the alternate ways to detect super-Chandrasekhar WD candidates. If~B-WDs are rotating with certain angular frequency, then they can efficiently emit gravitational radiation, provided that their magnetic field and rotation axes are not aligned~\cite{BG1996}, and~these gravitational waves can be detected by upcoming instruments. 
The dimensionless amplitudes of the two polarizations of a gravitational wave (GW) at a time $t$ are given by~\cite{BG1996,ZS1979}
\begin{eqnarray}
h_+ = h_0 {\rm sin}\,\chi \, \left[\frac{1}{2}{\rm cos}\,i\, {\rm sin}\,i \,{\rm cos}\,\chi \, {\rm cos}\,\Omega t - \frac{1+{\rm cos}^2 i}{2}{{\rm sin}\,\chi\,}{\rm cos}\,2\Omega t\right], \nonumber \\
h_{\times} = h_0 {\rm sin}\,\chi \left[\frac{1}{2}{\rm sin}\,i\, {\rm cos}\,\chi\, {\rm sin}\,\Omega t - {\rm cos}\,i\, {\rm sin}\,\chi\, {\rm sin}\,2\Omega t \right],
\end{eqnarray}
with $h_0 = (-6G/c^4)Q_{z^{\prime}z^{\prime}}(\Omega^2/d)$, where $Q_{z^{\prime}z^{\prime}}$ is the quadrupole moment of the distorted star, $\chi$ is the angle between the rotation axis $z^{\prime}$ and the body's third principal axis $z$, and $i$ is the angle between the rotation axis of the object and our line of sight. The~left panel of Figure~\ref{fig9} shows a schematic diagram of a pulsar with $z^{\prime}$ being the rotational axis and $z$ the magnetic field axis, where the angle between these two axes is $\chi$. The~GW amplitude is
\begin{equation}
h_0 = \frac{4G}{c^4} \frac{\Omega^2 \epsilon I_{xx}}{d},
\end{equation}
where $\epsilon = (I_{zz} - I_{xx})/I_{xx}$ is the ellipticity of the body, and $I_{xx}$, $I_{yy}$, $I_{zz}$ are the principal moments of inertia. We have used \emph{XNS} code~\cite{Pili2014} to simulate the underlying axisymmetric equilibrium configuration of B-WDs in general relativity. Moreover, we assume the distance between the WD and the detector to be 100~pc.

Since a pulsating WD can emit both dipole and GW radiations simultaneously, it is associated with both dipole and quadrupolar luminosities. Dipole luminosity for an axisymmetric WD is~\cite{Melatos2000}
\begin{equation}
L_D = \frac{B_p^2 R_p^6 \Omega^4}{2c^3} {\rm sin}^2 \, \chi \, F(x_0),
\end{equation}
where $x_0 = R_0 \Omega/c$, $B_p$ is the magnetic field strength at the pole, $R_p$ is the radius of the pole and $R_0$ is the average WD radius. The~function $F(x_0)$ is defined as
\begin{equation}
F(x_0) = \frac{x_0^4}{5(x_0^6 - 3x_0^4 + 36)} + \frac{1}{3(x_0^2 + 1)}.
\end{equation}

Similarly, the~quadrupolar GW luminosity is given by~\cite{ZS1979}
\begin{equation}
L_{GW} = \frac{2G}{5c^5}(I_{zz} - I_{xx})^2 \Omega^6 {\rm sin}^2\, \chi \, (1+15{\rm sin}^2 \, \chi).
\end{equation}

It should be noted that this formula is valid if $\chi$ is very small. The~total luminosity is due to both dipole and gravitational radiations. Therefore, 
$\Omega$ and $\chi$ decay with time due to both $L_D$ and $L_{GW}$, given by~\cite{Melatos2000}
\begin{eqnarray}
\nonumber
\frac{d(\Omega I_{z^{\prime}z^{\prime}})}{dt} = -\frac{2G}{5c^5}(I_{zz}-I_{xx})^2 \Omega^5 {\rm sin}^2\, \chi \, (1+15\, {\rm sin}^2\, \chi) - \frac{B_p^2 R_p^6 \Omega^3}{2c^3}{\rm sin}^2\, \chi \, F(x_0), \\
\\
\nonumber
I_{z^{\prime}z^{\prime}}\frac{d\chi}{dt} = -\frac{12G}{5c^5}(I_{zz} - I_{xx})^2 \Omega^4 {\rm sin}^3\, \chi\, {\rm cos}\, \chi - \frac{B_p^2 R_p^6 \Omega^2}{2c^3}{\rm sin}\, \chi \, {\rm cos}\, \chi \, F(x_0),\\
\end{eqnarray}
where $I_{z^{\prime}z^{\prime}}$ is the moment of inertia about the $z^{\prime}$-axis. Equations~(26) and (27) need to be solved simultaneously to obtain the timescale over which a WD can~radiate. 

The right panel of Figure~\ref{fig9} shows the dimensionless GW amplitudes for WDs as functions of their frequencies, along with the sensitivity curves of various detectors. We find that the isolated WDs may not be detected directly by LISA but~can be detected after integrating the signal-to-noise ratio (SNR) for 1 year. As~WDs are larger in size compared to NS, they cannot rotate as fast as NS and hence ground-based GW detectors such as LIGO, Virgo and KAGRA are not expected to detect the isolated WDs. 
These isolated WDs are also free from the noise due to the galactic binaries as well as from the extreme mass ratio inspirals (EMRIs). 

A pulsar-like object radiates GWs at two frequencies. When we observe such a GW signal whose strength remains unchanged during the observation time $T$, the~corresponding detector's cumulative SNR is given by~\cite{Jara1998,Bennett2010}
\begin{equation}
	{\rm SNR} = \sqrt{S/N_{\Omega}^2 + S/N_{2\Omega}^2}
\end{equation}
where
\begin{equation}
\langle S/N_{\Omega}^2 \rangle = \frac{{\rm sin}^2 \zeta}{100} \frac{h_0^2 T {\rm sin}^2 2\chi}{S_n(f)}, \langle S/N_{2\Omega}^2 \rangle = \frac{4{\rm sin}^2 \zeta}{25} \frac{h_0^2 T {\rm sin}^4 \chi}{S_n(2f)}
\end{equation}
where $\zeta$ is the angle between the interferometer arms and $S_n(f)$ is the detector's power spectral density (PSD) at the frequency $f$ with $\Omega = 2\pi f$. As~we mostly deal with space-based interferometers such as
LISA, we assume $\zeta = 60^{\circ}$. The~average is over all possible angles, including $i$, which determine the object's orientation with respect to the celestial sphere reference~frame.

\begin{figure}[H]
\begin{tabular}{cc}
\includegraphics[width=2.55in]{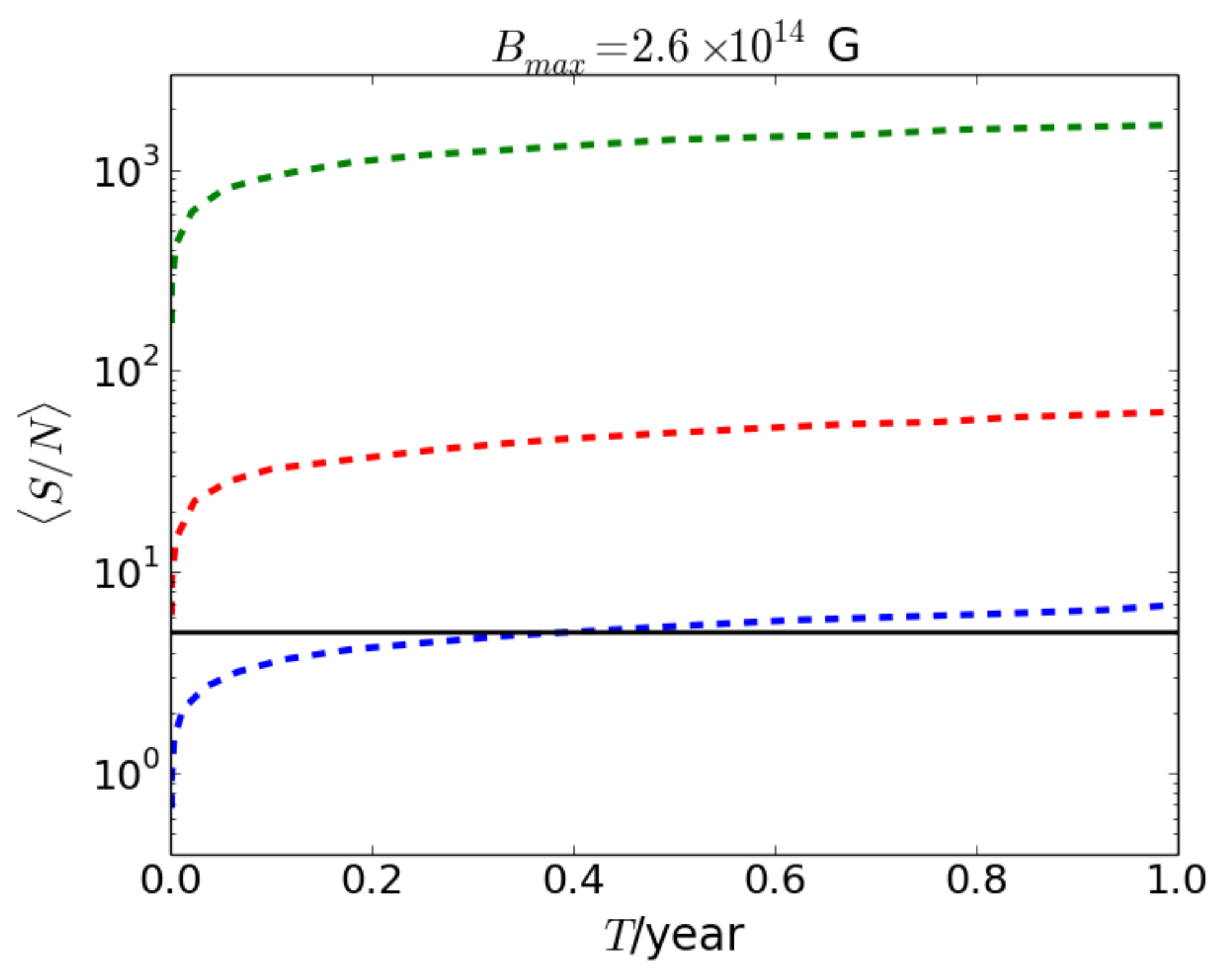} 
&\includegraphics[width=2.55in]{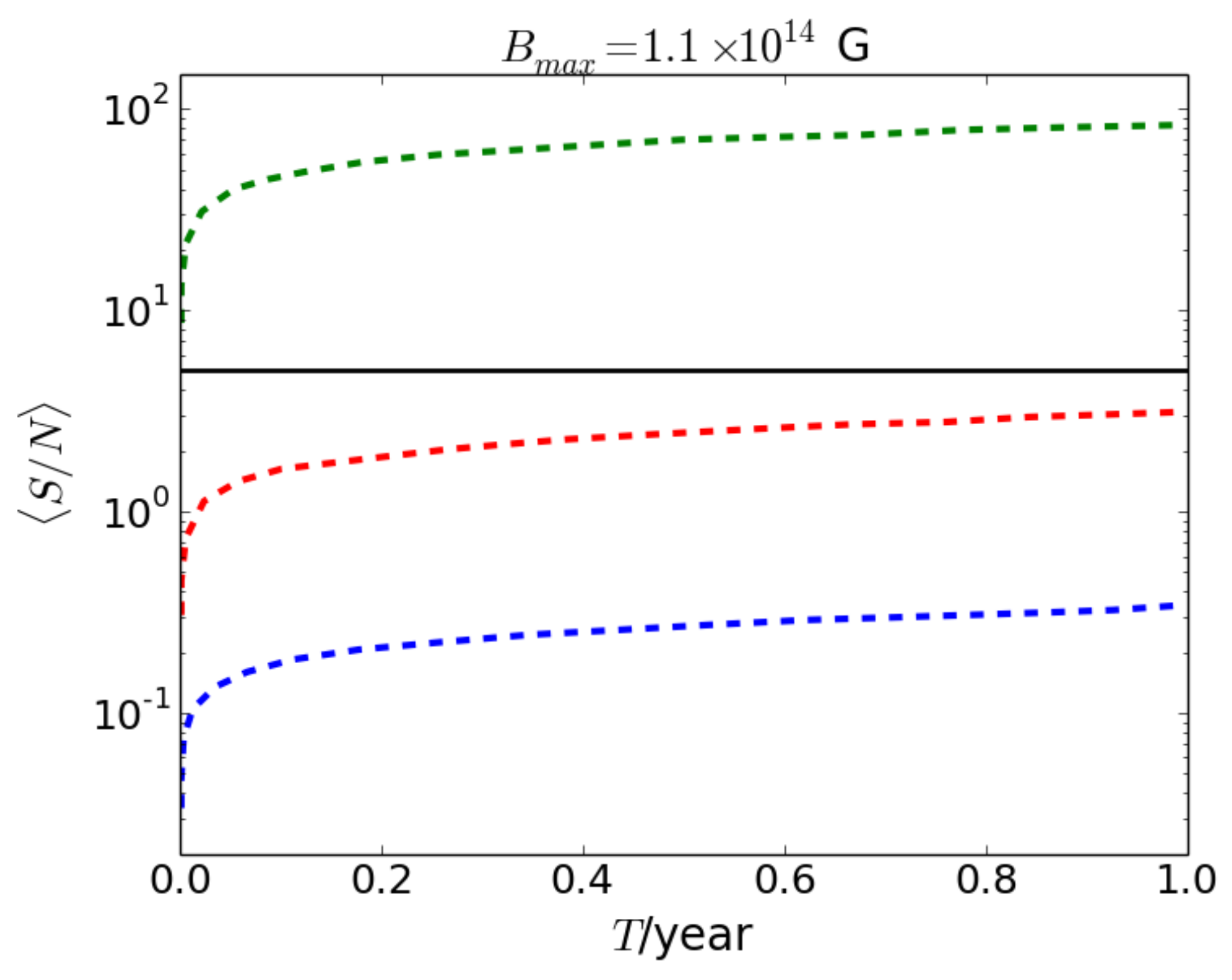}\\ 
\end{tabular}
	\caption{SNR as a function of integration time for a toroidal field-dominated WD with central density $\rho_c = 2\times10^{10}\, {\rm g\ cm^{-3}}$, spin period $P=2\, {\rm s}$, $\chi = 30^{\circ}$ and $d=100\, {\rm pc}$. The green (top) line represents \emph{ALIA}, the red (middle) line represents \emph{TianQin} and blue (bottom) line represents \emph{LISA}. The solid black line corresponds to $\langle S/N \rangle \approx 5$. See~\cite{Kalita2021}.}%
\label{fig10}
\end{figure}

Figure~\ref{fig10} shows the SNR as a function of time for toroidal field-dominated WDs with different field strengths. We assume that these toroidal-dominated WDs have a poloidal surface field which is nearly four orders of magnitude smaller than the maximum toroidal field $B_{\rm max}$ inside the WD. Of~course, such a poloidal field cannot change the shape and size of the WD, as does the toroidal field. The~surface field strength is relatively very small (as is the dipole luminosity), and so it hardly changes $\Omega$ and $\chi$ within a 1 yr period. The~left panel of Figure~\ref{fig10} shows the SNR for a B-WD with $B_{\rm max} = 2.6\times10^{14}\, {\rm G}$ with mass $1.7 M_{\odot}$. All the GW detectors except \emph{LISA} can easily detect such a WD almost immediately, and \emph{LISA} can detect it in 5 months of integration. In~contrast, when the field strength decreases ($B_{\rm max} \approx 10^{14}\, {\rm G}$) the SNR decreases, and \emph{LISA} and \emph{TianQin} can no longer detect them, as shown in the right panel. However, they can still be detected by \emph{ALIA}, \emph{BBO} and \emph{DECIGO} within 1 yr of integration~time.

\begin{figure}[H]
\begin{tabular}{cc}
\includegraphics[width=2.55in]{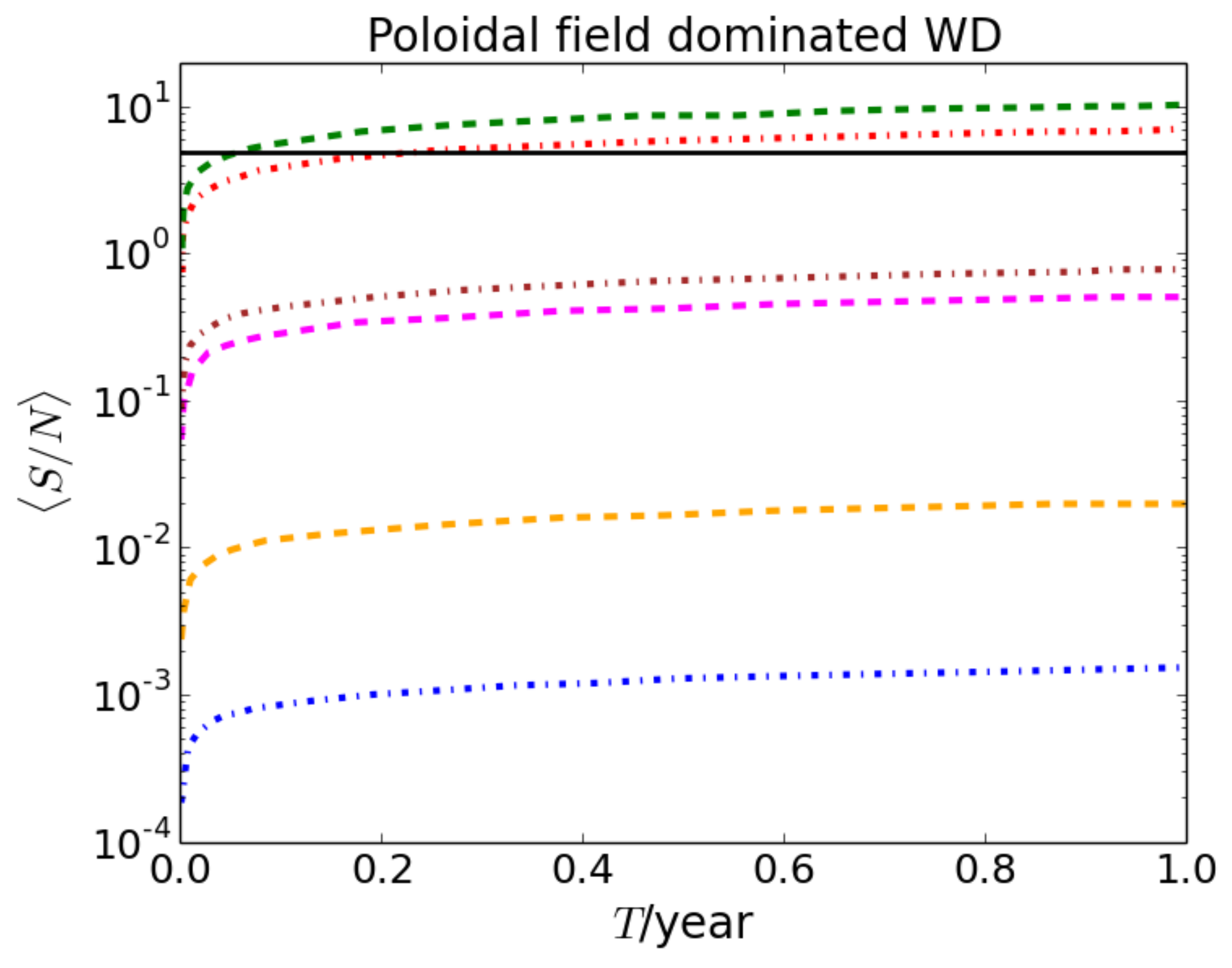} 
&\includegraphics[width=2.55in]{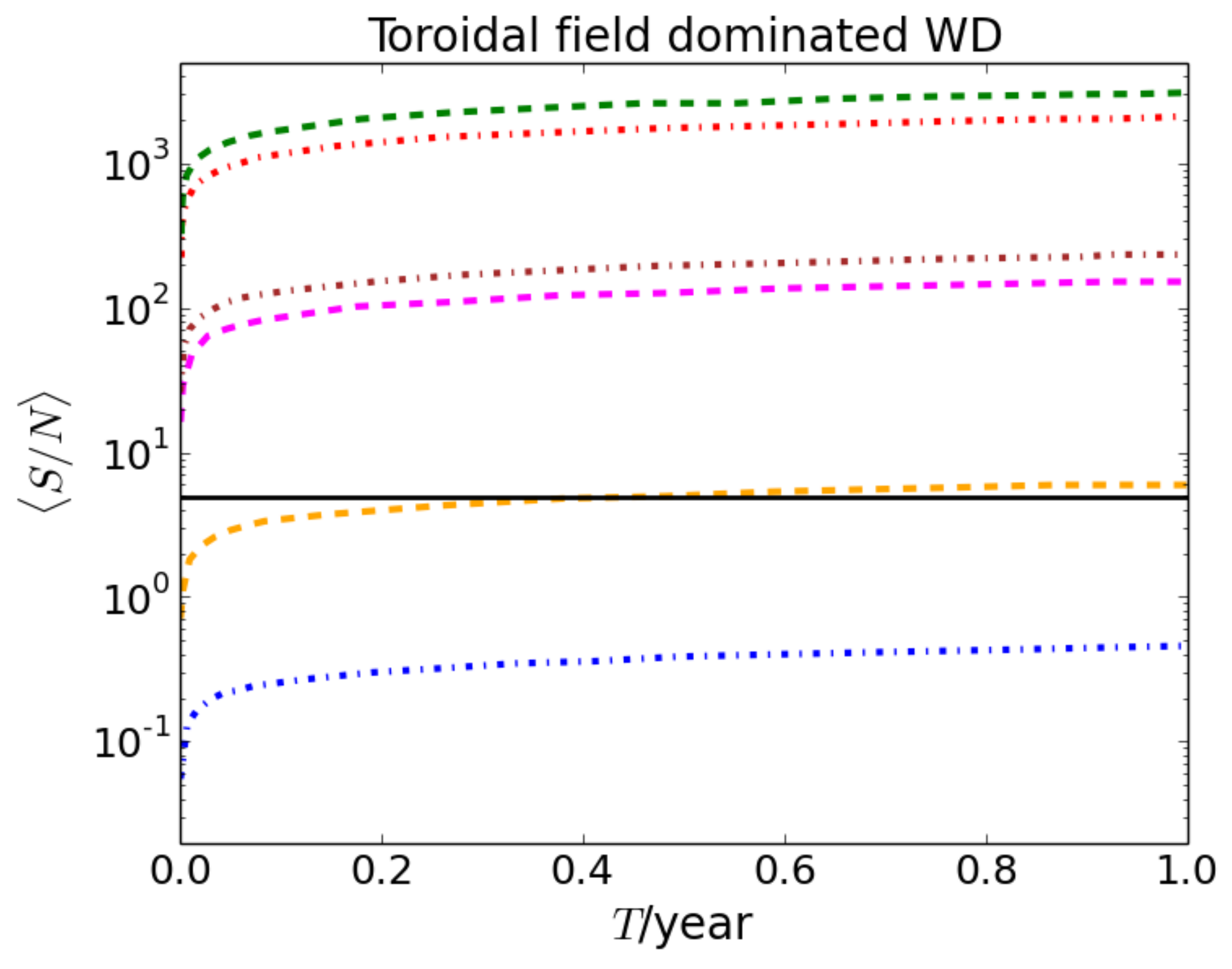}\\ 
\end{tabular}
	\caption{SNR as a function of integration time for XTE J1810-197 assuming $\chi=45^{\circ}$.  The first, second and fifth lines from top correspond to a WD with radius 3000 km for \emph{BBO}, \emph{DECIGO} and \emph{ALIA} respectively, while third, fourth and sixth lines from top represent a WD with radius 1000 km for
	\emph{BBO}, \emph{DECIGO} and \emph{ALIA} respectively. The solid black line corresponds to $\langle S/N \rangle \approx 5$. See~\cite{Kalita2021}.}%
\label{fig11}
\end{figure}

Figure~\ref{fig11} shows the SNR as a function of integration time for XTE J1810-197, assuming $\chi=45^{\circ}$. We choose the maximum radius of this source to be 3000 km instead of 4000 km if it is a WD because its spin is fast and the \emph{XNS} code does not run for 4000 km with such high rotation frequency. The~left panel of Figure~\ref{fig11} shows that \emph{BBO} and \emph{DECIGO} would be able to detect it within 20 d and 100 d, respectively, if~it is a 3000 km poloidal field-dominated WD. If~it is a toroidally dominated WD, \emph{BBO} and \emph{DECIGO} could immediately detect it, and \emph{ALIA} would be able to detect it within 5 months only if it is a 3000 km toroidally dominated WD (see right panel). 

There are many other plausible detectabilities, e.g.,~via their activity in 
binary systems, which will be explored in the future. 

\section{Matter Anisotropy Effects in Highly Magnetised White~Dwarfs}
\label{Sec8}
It has been shown that the presence of a strong magnetic field, the~anisotropy of dense matter, and~the orientation of a magnetic field can significantly influence the properties of neutron and quark stars~\cite{Deb2021}. The stability of them
and B-WDs is not achieved unless one considers the anisotropy of the system arising from the combined effects of~\cite{Deb2021,Deb2022}: (i)~anisotropy due to strong magnetic fields and (ii) anisotropy of the system~fluid.

The effective contributions from the matter and magnetic field lead to the system density given by
\begin{equation}
\tilde{\rho} = \rho + \frac{B^2}{8\pi}.
\end{equation}

The system pressure along the direction of magnetic field is represented as parallel pressure and takes the form based on magnetic field orientations as
\begin{equation}
p_{||} = \begin{cases}
p_r - \frac{B^2}{8\pi}, {\rm \ for\ radial\ orientation}\\
p_t - \frac{B^2}{8\pi}, {\rm \ for\ transverse\ orientation}.
\end{cases}
\end{equation}

Similarly, the~system pressure that aligns perpendicular to the magnetic fields is defined as transverse pressure and is based on the magnetic field orientations
\begin{equation}
p_{\perp} = \begin{cases}
p_t + \frac{B^2}{8\pi}, {\rm \ for\ radial\ orientation}\\
p_r + \frac{B^2}{8\pi}, {\rm \ for\ transverse\ orientation}.
\end{cases}
\end{equation}

The essential magnetostatic stellar equations which describe static, spherically symmetric B-WDs are
\begin{equation}
\frac{dm}{dr} = 4\pi \left(\rho + \frac{B^2}{8\pi}\right)r^2,
\end{equation}
and for radial orientation (RO)
\begin{equation}
\frac{dp_r}{dr} = \frac{-(\rho+p_r)\frac{4\pi r^3 \left(p_r - \frac{B^2}{8\pi}\right)+m}{r(r-2m)} + \frac{2}{r}\Delta}{\left[1- \frac{d}{d\rho}\left(\frac{B^2}{8\pi}\right)\frac{d\rho}{dp_r}\right]},
\end{equation}
whereas for transverse orientation (TO)
\begin{equation}
\frac{dp_r}{dr} = \frac{-\left(\rho+p_r+\frac{B^2}{4\pi}\right)\frac{4\pi r^3 \left(p_r + \frac{B^2}{8\pi}\right)+m}{r(r-2m)} + \frac{2}{r}\Delta}{\left[1+ \frac{d}{d\rho}\left(\frac{B^2}{8\pi}\right)\frac{d\rho}{dp_r}\right]}.
\end{equation}

We describe the effective anisotropy of the stars with $\Delta$, which depends on the magnetic field orientations given by $p_t - p_r + \frac{B^2}{4\pi}$ in case of RO and $p_t - p_r - \frac{B^2}{4\pi}$ for TO. We have for~RO
\begin{equation}
\Delta = \kappa \frac{(\rho+\rho_r)\left(\rho+3p_r - \frac{B^2}{4\pi}\right)}{\left(1 - \frac{2m}{r}\right)}r^2,
\end{equation}
and for TO
\begin{equation}
\Delta = \kappa \frac{(\rho+\rho_r+\frac{B^2}{4\pi})\left(\rho+3p_r + \frac{B^2}{2\pi}\right)}{\left(1 - \frac{2m}{r}\right)}r^2,
\end{equation}
where $\kappa$ is a dimensionless constant that describes the strength of anisotropy within the stellar structure. Note that $\kappa = 0$ implies the anisotropy effects that arise due to matter properties
and magnetic field both vanish. However, the~case of $B = 0$ but $\kappa \neq 0$ implies that only the anisotropy due to magnetic field vanishes. We have shown that highly magnetized WD models, which do not account for the combined anisotropic effects of the fluid and field, are eliminated, as they suffer from an
instability at the stellar~center. 

To solve the magneto-hydrostatic stellar structure equations from the stellar center to surface, it is needed to supply an EoS along with a functional form of $\Delta$. Here, we consider the EoS proposed by Chandrasekhar to describe 
degenrate electrons of WDs as
\begin{equation}
p_r = \frac{\pi m_e^4 c^5}{3h^3} \left[x(2x^2 - 3) \sqrt{x^2 + 1} + 3 {\rm sinh}^{-1} x\right],\ \rho = \frac{8\pi \mu_e m_H (m_e c)^3}{3h^3} x^3,
\end{equation}
where $m_e$ is the mass of an electron, $m_H$ is the mass of a hydrogen atom, $h$ is Planck's constant, $\mu_e$ is the mean molecular weight per electron, and~$x = p_F/m_{e}c$ with $p_F$ is the Fermi momentum. For~the carbon-oxygen WDs, $\mu_e=2$.

\begin{figure}[H]
\centering 
\begin{tabular}{ccc}
\includegraphics[width=2.2in]{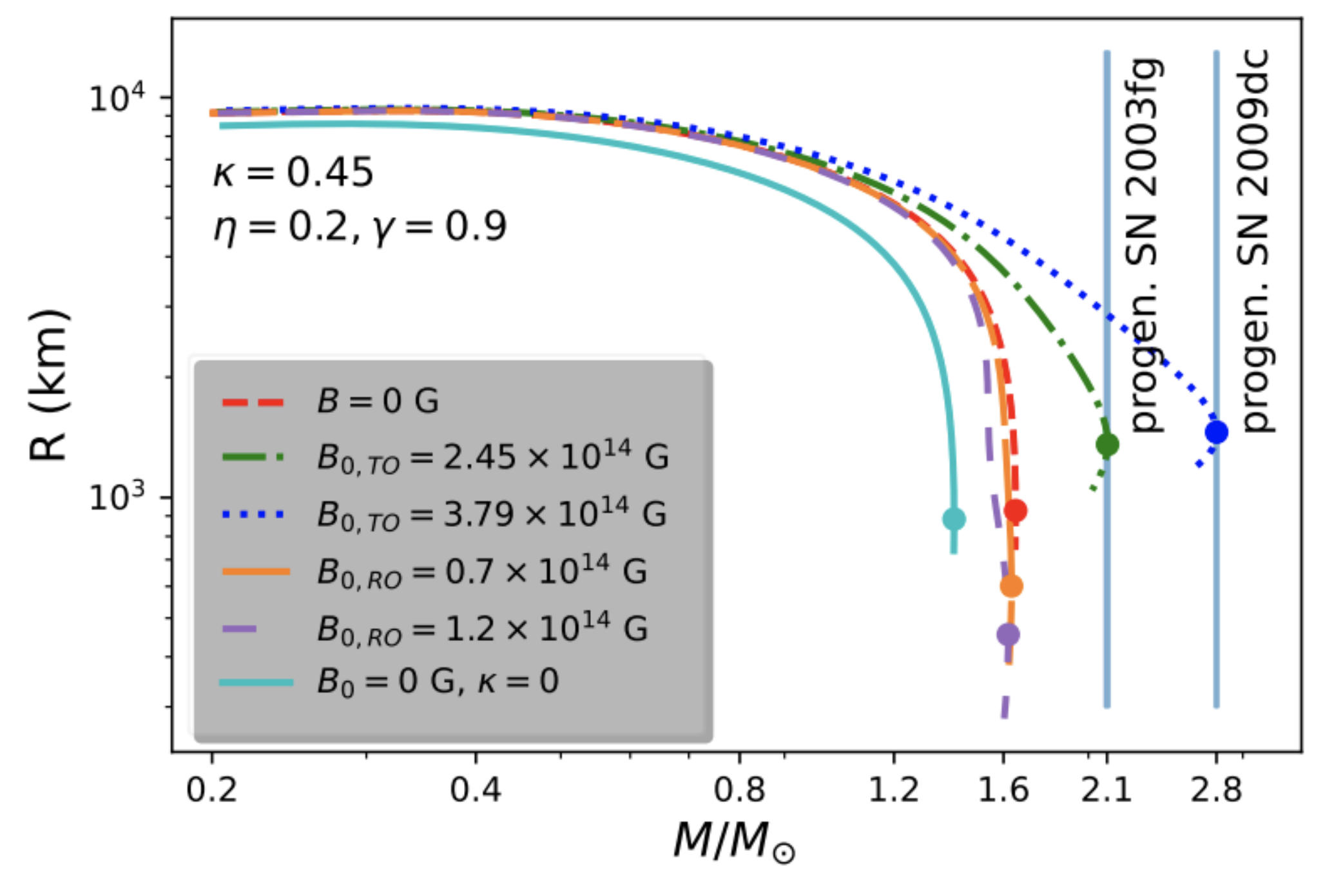}
&\includegraphics[width=2.28in]{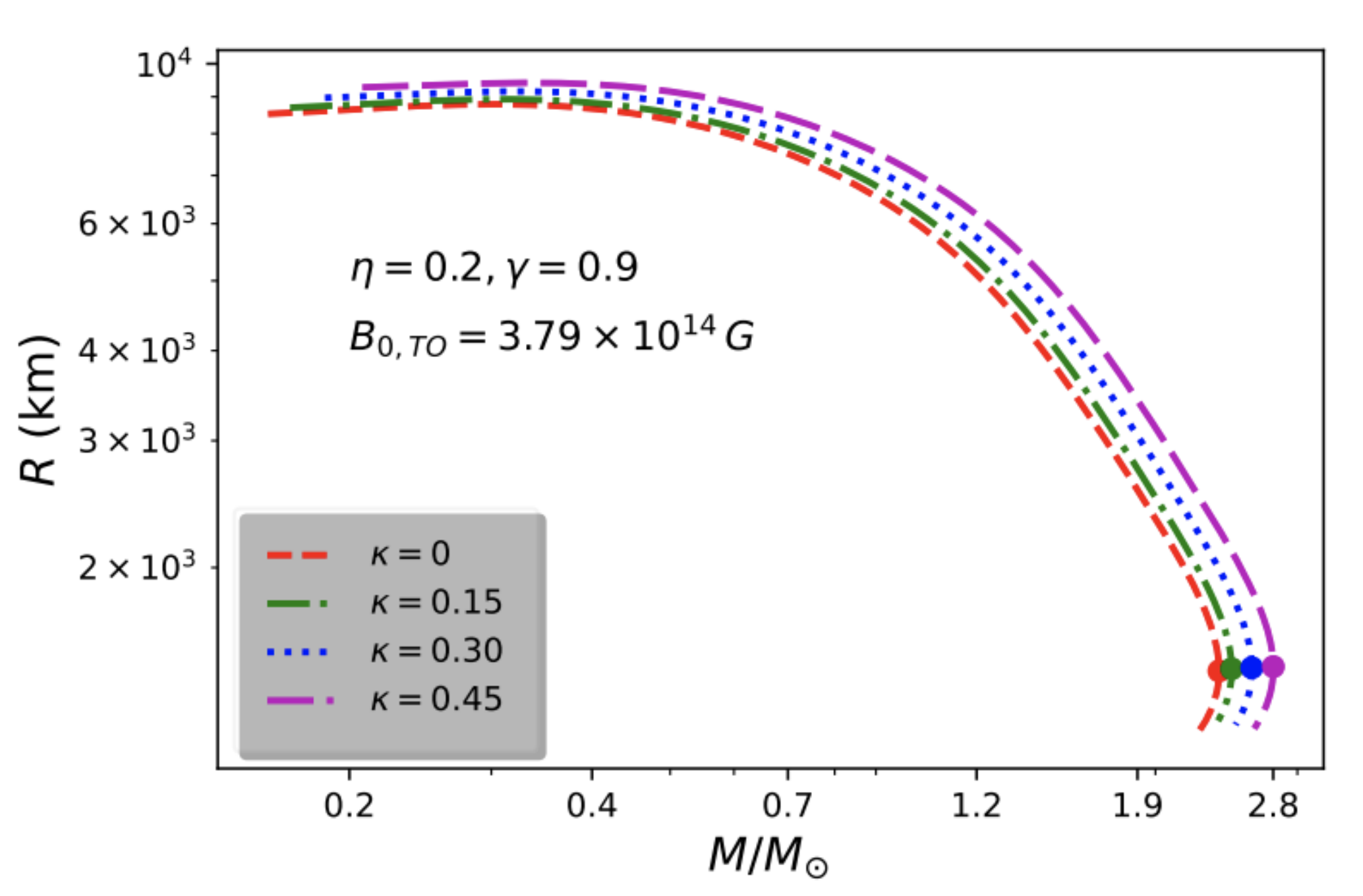}
&\includegraphics[width=2.2in]{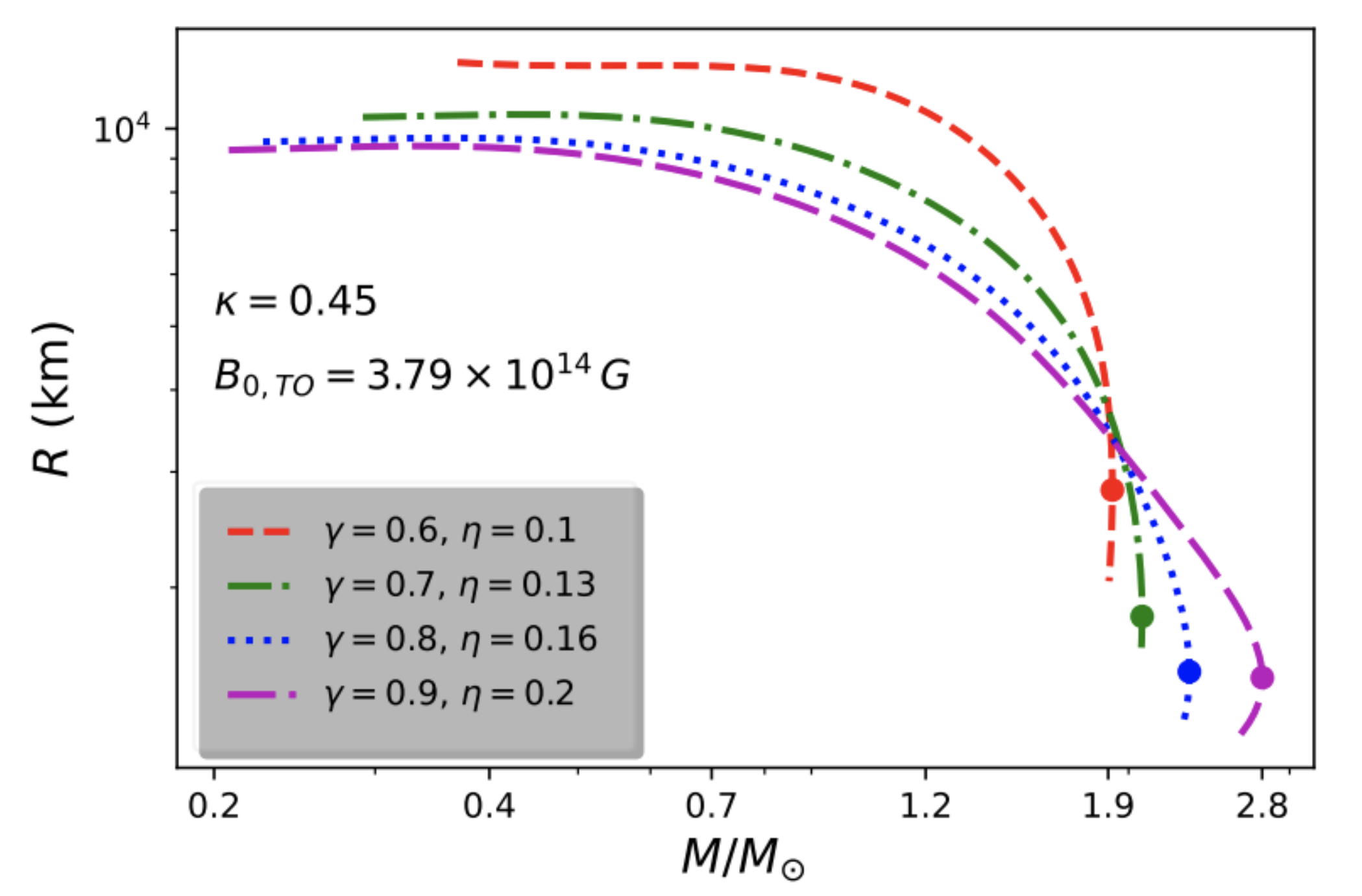}\\
({\bf a})&({\bf b})&({\bf c})\\
\end{tabular}
	\caption{Stellar radius ($R$) as a function of gravitational mass ($M/M_{\odot}$) for different (\textbf{a}) $B_0$ (left~panel), (\textbf{b}) $\kappa$ (middle panel), and~(\textbf{c}) $\eta$ and $\gamma$ (right~panel). Solid circles represent the stars with the maximum-possible masses. See~\cite{Deb2022}.}%
\label{fig12}
\end{figure}

We show the mass-radius relations of B-WDs for different $B_0$, $\kappa$ and $\gamma$ in Figure~\ref{fig12}. For~TO fields with $B_0 = 3.79\times10^{14}\, {\rm G}$, a~maximum mass B-WD of $2.8 M_{\odot}$ is obtained, whose radius is 1457.67 km. For an~RO field with $B_0 = 1.2\times10^{14}\, {\rm G}$, the~maximum mass drops to $1.62 M_{\odot}$, and the radius of the B-WD is 454.67 km. For~$B_{0,TO} = 3.79\times10^{14}\, {\rm G}$, the~maximum mass and the corresponding radius of B-WDs increase by $\sim 70\%$ and $\sim 57\%$, respectively, compared to the non-magnetized but anisotropic case. For~$B_{0,RO} = 1.2\times10^{14}\, {\rm G}$, the~maximum mass and the corresponding radius decrease by $\sim 2\%$ and $\sim 52\%$, respectively, compared to the values
of non-magnetized but anisotropic WDs. Even without considering the magnetic field and incorporating the effects of local anisotropy due to the fluid, it is possible to push the maximum mass of WDs beyond the Chandrasekhar mass limit. For~example, by~considering $\kappa = 2/3$, we obtain a maximum mass for a non-magnetized but anisotropic WD of $1.81 M_{\odot}$. The~corresponding radius is 956.08 km. These values are $\sim 29\%$ and $\sim 8\%$, respectively, higher than the respective values of WDs at the Chandrasekhar mass limit. Moreover, with~increasing $\eta$ and $\gamma$ for the TO field case, the~mass of anisotropic B-WDs changes significantly, as~can be seen in the right panel of Figure~\ref{fig12}. It shows that the maximum mass of a B-WD with $\gamma = 0.9$ and $\eta = 0.2$ increases by $\sim 46\%$ compared to $\gamma = 0.6$ and $\eta = 0.1$.

\vspace{-5pt}

\begin{figure}[H]
\centering 
\begin{tabular}{ccc}
\includegraphics[width=2.232in]{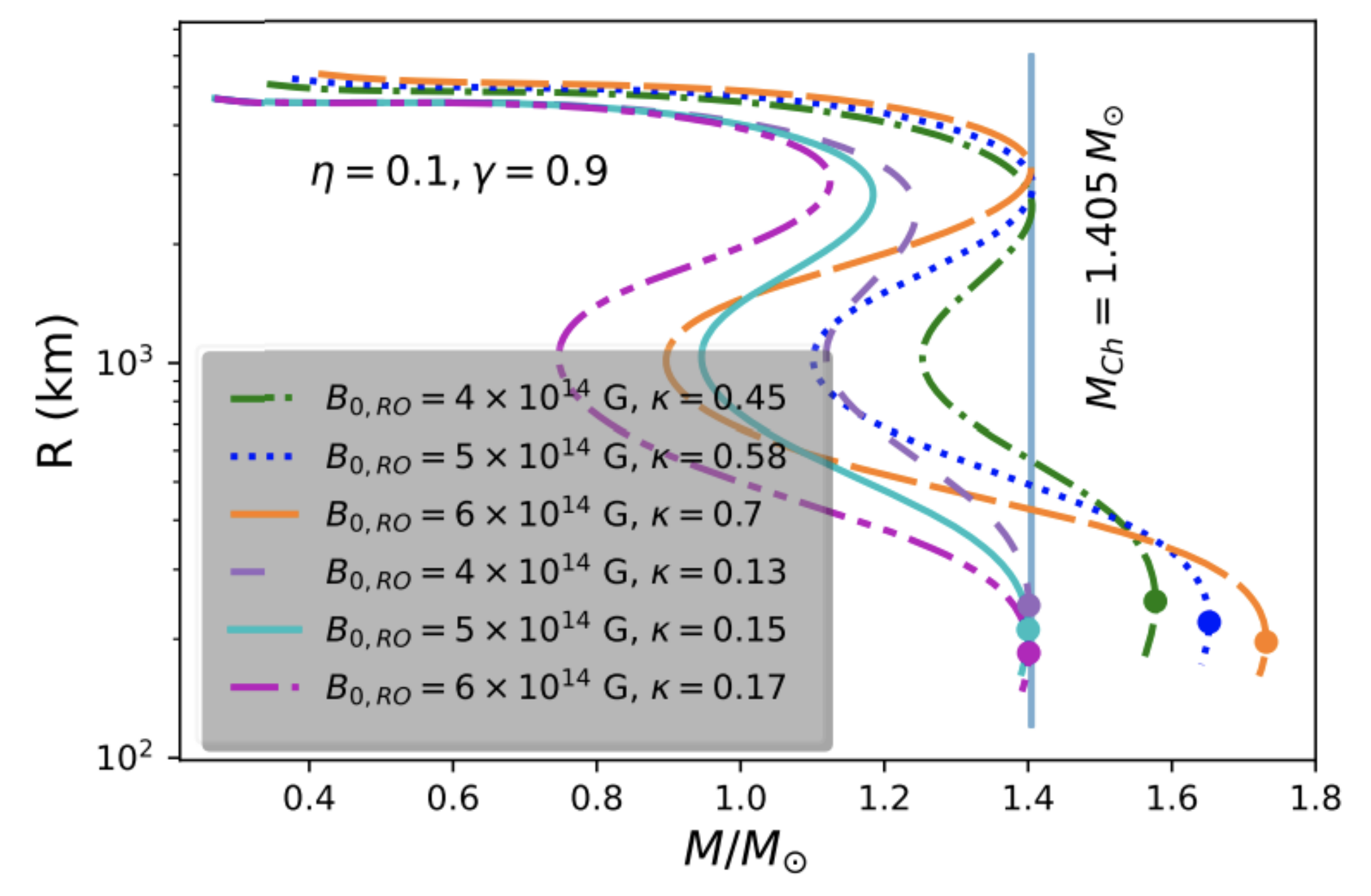}
&\includegraphics[width=2.23in]{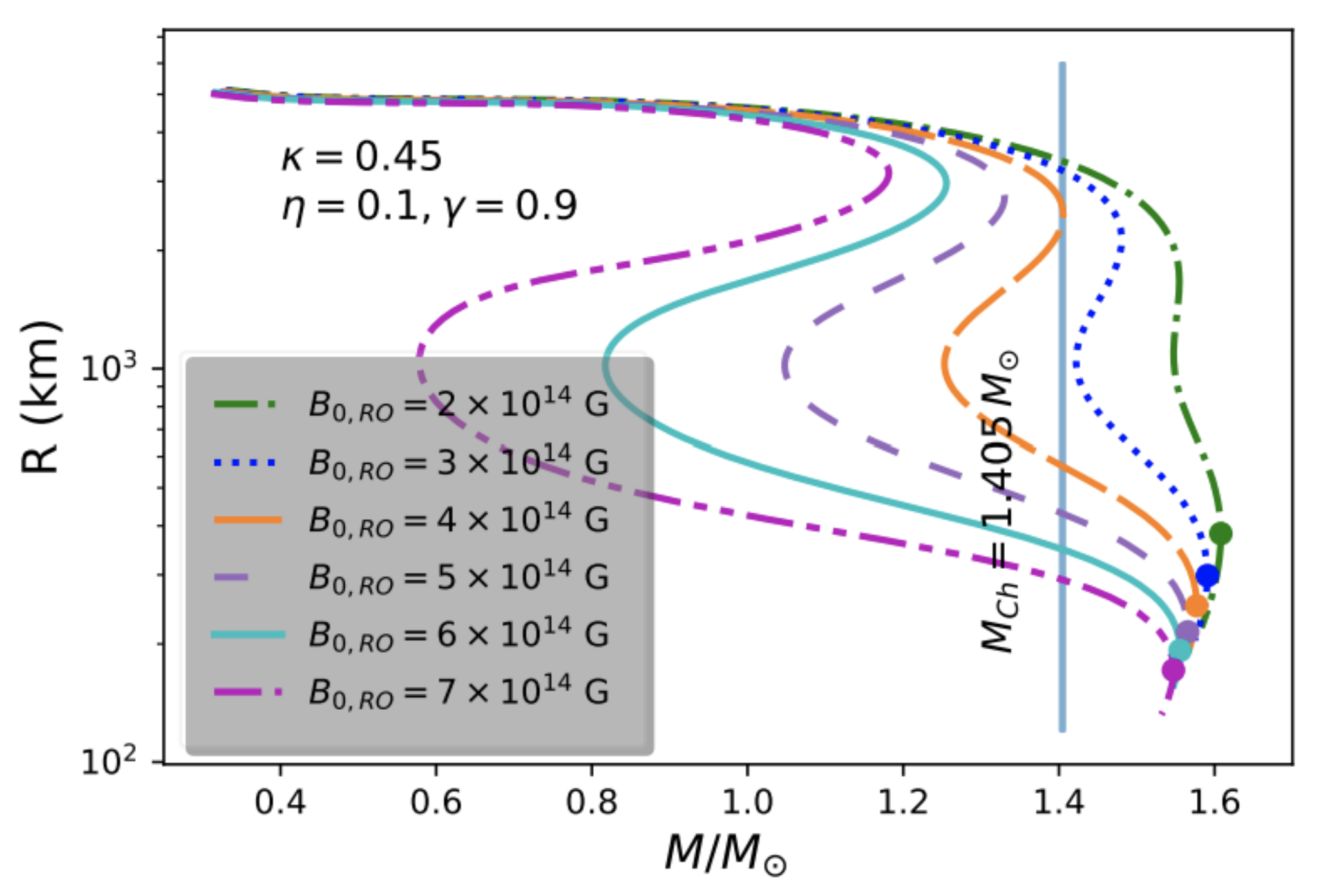}
&\includegraphics[width=2.23in]{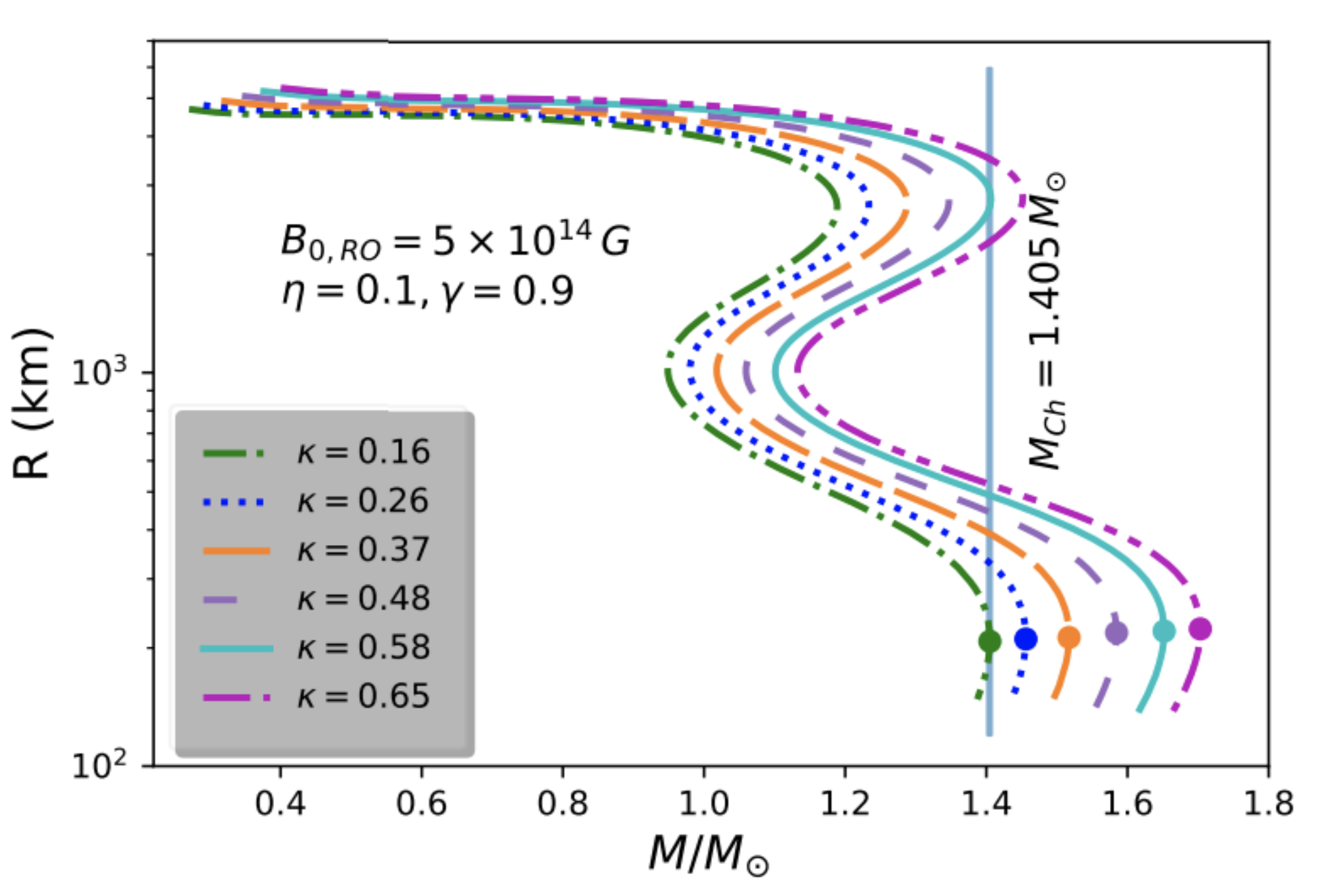}\\
({\bf a})&({\bf b})&({\bf c})\\
\end{tabular}
	\caption{Stellar radius ($R$) as a function of gravitational mass ($M/M_{\odot}$) for varying (\textbf{a}) $B_{0,RO}$ and $\kappa$ (left panel), (\textbf{b}) $B_{0,RO}$ (middle panel), and~(\textbf{c}) $\kappa$ (right panel). Solid circles represent the stars with the maximum possible masses. See~\cite{Deb2022}.}%
\label{fig13}
\end{figure}

For RO fields, we can explain sub-, standard- and super-Chandrasekhar B-WDs by making appropriate choices of $B_{0,RO}$ and $\kappa$ (see Figure~\ref{fig13}). By~changing both $B_{0,RO}$ and $\kappa$, as~shown in the left panel of Figure~\ref{fig13}, we successfully explain both (i) the sub- and standard-Chandrasekhar progenitor B-WDs and (ii) the standard and super-Chandrasekhar progenitor
B-WDs by~a single mass-radius curve for the respective cases. On~the other hand, through changes of only $B_{0,RO}$ or $\kappa$, sub-, standard- and super-Chandrasekhar B-WDs line up in a series of mass-radius curves; see, e.g.,~the middle and right panels of Figure~\ref{fig13}. This leads to a complete explanation of under-, regular- and over-luminous SNeIa in a single~theory. 

\vspace{-5pt}

\section{Conclusions}
\label{Sec9}
Reviewing the work in last decade or so, we have confirmed that, at~least theoretically, massive NSs and WDs, heavier
than their conventional counterparts as observed/inferred from recent data,
are possibly highly magnetized, rotating stable compact stars. They have 
multiple implications including enigmatic peculiar over-luminous SNeIa.
Numerical simulations based on Cambridge stellar evolution code STARS argue
B-WDs to be toroidally (centrally) dominated with lower, maybe dipole,
surface magnetic fields. The~new generic mass-limit of WDs seems to be 
well above $2M_\odot$, around $2.8M_\odot$, depending on the magnetic field
profile/geometry and rotational~frequency. 

However, these massive compact objects, particularly WDs, have hardly been 
observed directly due to their apparently very low luminosity. On~the other
hand, the~presence of magnetic fields and rotation by definition brings in 
anisotropy in these compact objects, leading them to be triaxial when the 
magnetic and rotation axes are misaligned. Therefore, they are expected to 
emit continuous GW. Hence, B-WDs can be detected directly by the missions
LISA (in one year integration time), DECIGO/BBO and the magnetised NSs
can be detected by aLIGO, aVIRGO, Einstein~Telescope.

Nevertheless, the~magnetic fields therein start decaying beyond million years,
hence these massive compact objects may not survive beyond this time.
Moreover, due to their electromagnetic and gravitational radiation, the~angular velocity and the misalignment angle between magnetic and
rotation axes decrease, leading them to lose their pulsar nature and detection 
possibility. This is another reason why they are so rare and/or hard to
detect unless targeted at an appropriate time. Therefore, appropriate missions
in GW astronomy and otherwise, e.g.,~radio astronomy, should be planned to
probe~them.

\vspace{6pt}

\authorcontributions{\textls[-20]{Both the authors share an equal authorship.
However, various figures presented in this manuscript are based on previous
works by B.M. and his collaborators. 
The work was conceptualized by B.M.;
the methodology was initiated by B.M. and further was carried out 
by B.M. and M.B.; investigation was initiated by B.M. and was further 
taken over by M.B.; the first draft was prepared by M.B. and further was
revised and edited by B.M. and M.B.}}

\funding{M.B. would like to acknowledge support from the Eberly Research Fellowship at the Pennsylvania State University.} 

\acknowledgments{B.M. thanks the organizers of ``The Modern Physics of Compact Stars and Relativistic Gravity 2021" held in September 27-30, 2021, in  
Yerevan, Armenia, to invite for a talk, based on which the present paper
is composed. }
  
  \conflictsofinterest{The authors declare no conflict of interest.}

\appendixtitles{yes} 
\appendixsections{multiple} 

\reftitle{References}

\end{document}